\newif\ifMARKUP
\MARKUPtrue

\ifMARKUP
\documentclass[prx,notitlepage,twocolumn,superscriptaddress,longbibliography]{revtex4-1}

\else
\documentclass[prx,notitlepage,twocolumn,superscriptaddress,longbibliography,linenumbers]{revtex4-1}

\fi

\usepackage[english]{babel}
\usepackage{amsmath,amssymb,bbm,mathrsfs,mathtools,braket,color,%
  graphicx,comment,textcomp,amsfonts,dsfont,lettrine,units}
\usepackage[colorlinks,citecolor=blue,urlcolor=blue]{hyperref}
\newcommand{\Z}{\mathbb{Z}}
\newcommand{\C}{\mathbb{C}}
\newcommand{\R}{\mathbb{R}}

\newcommand{\e}{\mathrm{e}}
\renewcommand{\i}{\mathrm{i}}

\newcommand{\abs}[1]{\left\lvert #1 \right\rvert}
\newcommand{\norm}[1]{\left\lVert #1 \right\rVert}
\newcommand{\tr}{\mathop{\mathrm{tr}}}

\renewcommand{\Re}{\mathop{\mathrm{Re}}}
\renewcommand{\Im}{\mathop{\mathrm{Im}}}

\newcommand{\sect}[1]{~\\\noindent{\large{\bf{#1}}~\\ \noindent}}
\newcommand{\subsect}[1]{\noindent{\bf{#1.}}}

\definecolor{mygreen}{rgb}{0,0.5,0}

\begin{document}

\title{Digital Quantum Simulation, Trotter Errors, and Quantum Chaos of the
  Kicked Top}

\author{Lukas M. Sieberer}

\email{lukas.sieberer@uibk.ac.at}

\affiliation{Center for Quantum Physics, University of Innsbruck, 6020
  Innsbruck, Austria}

\affiliation{Institute for Quantum Optics and Quantum Information of the
  Austrian Academy of Sciences, A-6020 Innsbruck, Austria}

\author{Tobias Olsacher}

\affiliation{Center for Quantum Physics, University of Innsbruck, 6020
  Innsbruck, Austria}

\author{Andreas Elben}

\affiliation{Center for Quantum Physics, University of Innsbruck, 6020
  Innsbruck, Austria}

\affiliation{Institute for Quantum Optics and Quantum Information of the
  Austrian Academy of Sciences, A-6020 Innsbruck, Austria}

\author{Markus Heyl}

\affiliation{Max Planck Institute for the Physics of Complex Systems,
  N\"othnitzer Str.~38, 01187 Dresden, Germany}

\author{Philipp Hauke}

\affiliation{Kirchhoff-Institute for Physics, Heidelberg University, 69120
  Heidelberg, Germany}

\affiliation{Institute for Theoretical Physics, Heidelberg University, 69120
  Heidelberg, Germany}

\author{Fritz Haake}

\affiliation{Faculty of Physics, University of Duisburg-Essen, 47048 Duisburg, Germany}

\author{Peter Zoller}

\affiliation{Center for Quantum Physics, University of Innsbruck, 6020
  Innsbruck, Austria}

\affiliation{Institute for Quantum Optics and Quantum Information of the
  Austrian Academy of Sciences, A-6020 Innsbruck, Austria}

\date{\today}

\begin{abstract}
  This work aims at giving Trotter errors in digital quantum simulation (DQS) of
  collective spin systems an interpretation in terms of quantum chaos of the
  kicked top. In particular, for DQS of such systems, regular dynamics of the
  kicked top ensures convergence of the Trotterized time evolution, while chaos
  in the top, which sets in above a sharp threshold value of the Trotter step
  size, corresponds to the proliferation of Trotter errors. We show the
  possibility to analyze this phenomenology in a wide variety of experimental
  realizations of the kicked top, ranging from single atomic spins to
  trapped-ion quantum simulators which implement DQS of all-to-all interacting
  spin-\nicefrac{1}{2} systems. These platforms thus enable in-depth studies of
  Trotter errors and their relation to signatures of quantum chaos, including
  the growth of out-of-time-ordered correlators.
\end{abstract}

\maketitle

In digital quantum simulation (DQS), unitary Hamiltonian evolution is decomposed
into a sequence of quantum gates. A common approach to achieve this
decomposition utilizes Suzuki-Trotter formulas~\cite{Trotter1959, Suzuki1976} to
approximately factorize the time evolution operator~\cite{Lloyd1073, Lanyon57,
  Peng2005, Barends2016, Langford2016, OMalley2016, Barends2015, Barreiro2011,
  Martinez2016, Salathe2015}. It is a fundamental conceptual question under
which conditions this ``Trotterization'' is a controlled approximation. A recent
numerical study~\cite{Heyl2019} by some of us found that Trotter errors in DQS
of generic many-body systems remain bounded below and proliferate above a
dynamical transition to \emph{many-body quantum
  chaos}~\cite{DAlessio2016}. Motivated by these findings we revisit the kicked
top, a paradigmatic model of \emph{single-body quantum
  chaos}~\cite{Haake2010}. Resorting to this well-studied model system allows us
to gain insights into Trotter errors in DQS of collective spin
systems. Moreover, the kicked top connects smoothly to a paradigmatic DQS of an
Ising chain with long-ranged power-law interactions.
  
The dynamics of the kicked top is described by the time-dependent Hamiltonian
\begin{equation}
  \label{eq:H-kicked-top}
  H(t) = H_x + \tau H_z \sum_{n \in \Z} \delta(t - n \tau),
\end{equation}
which combines precession of the spin of the top $S$ around a fixed axis,
$H_x = h_x S_x$, with non-linear ``kicks'' given by $H_z = J_z S_z^2/(2 S + 1)$,
which are applied periodically at times $t = n \tau$ for all $n \in \Z$. Here,
$S_{\mu}$ with $\mu = x, y, z$ are quantum angular momentum operators. The
inverse spin size can be regarded as an effective Planck constant,
$\hbar_{\mathrm{eff}} = 1/S$~\cite{Haake2010}, and in the limit $S \to \infty$,
a semiclassical description of the dynamics applies. Then, the precession of a
\emph{classical} top is only slightly perturbed by weak kicks, whereas strong
kicks cause the top to tumble chaotically~\cite{Haake2010}. This classical
chaotic motion is reflected in the spectrum of the Floquet operator (we set the
``true'' Planck constant to unity, $\hbar = 1$),
\begin{equation}
  \label{eq:U-kicked-top}
  U_{\tau} = \e^{-\i H_z \tau} \e^{-\i H_x \tau},
\end{equation}
which determines the evolution of a \emph{quantum} kicked top during a period of
duration $\tau$: In the chaotic regime of strong kicking, the spectral
statistics of $U_{\tau}$ is described by Dyson's ensemble of random orthogonal
matrices~\cite{Haake2010}. Indeed, it is a defining feature of quantum chaotic
systems that their spectral statistics are universal and obey predictions from
random-matrix theory (RMT)~\cite{Haake2010}. Among the model systems of quantum
chaos, the kicked top stands out due to its extraordinary faithfulness to RMT.
The discovery of this property initiated a surge of theoretical
studies
. Recent developments~\cite{Bandyopadhyay2015, Bhosale2017, Bhosale2018,
  Kumari2018} include proposals~\cite{Swingle2016} to diagnose chaos in the
kicked top by measuring out-of-time-ordered correlators, and the discovery of
critical quasienergy states~\cite{Bastidas2014}. Experimentally, the kicked top
was realized as the spin of single Caesium atoms~\cite{Chaudhury2009} with
$S = 3$, as the collective spin of an ensemble of three superconducting
qubits~\cite{Neill2016} corresponding to $S = 3/2$, and very recently in NMR as
a spin $S = 1$ composed of two spin-\nicefrac{1}{2}
nuclei~\cite{Krithika2019}. Novel experimental possibilities include atomic
species with high spin becoming available in labs such as
Dysprosium~\cite{Chalopin2018} with $S = 8$ and Erbium~\cite{Baier2018} with
$S = 19/2$, and also the spin $S = 7/2$ of $^{123}\mathrm{Sb}$
nuclei~\cite{Mourik2018}; implementations as collective spin in quantum
simulators of all-to-all coupled spin-\nicefrac{1}{2} with ``flip-flop''
qubits~\cite{Tosi2017}, or trapped ions~\cite{Britton2012, Garttner2017} in
which $S \gtrsim 50$ can be realized; and condensates of ultracold bosonic
atoms~\cite{Strobel2014} corresponding to even larger collective spins $S$ on
the order of several hundreds.

Formally, the dynamics of the kicked top, described by repeated application of
the Floquet operator $U_{\tau}$ in Eq.~\eqref{eq:U-kicked-top}, is equivalent to
DQS of a system with Hamiltonian $H = H_x + H_z$: In DQS, time evolution is
often ``Trotterized''~\cite{Trotter1959, Suzuki1976}, i.e., the run-time $t$ of
a simulation is split into $n$ steps of duration $t/n$, and within each step the
time evolution operator is approximately factorized,
\begin{equation}
  \label{eq:Suzuki-Trotter}
  U(t) = \e^{-\i H t} \approx \left( \e^{-\i H_z t/n} \e^{-\i H_x t/n} \right)^n =
  U_{\tau}^n.
\end{equation}
We note that other ways of decomposing the time evolution operator are being
studied in the literature~\cite{Childs2012, Berry2015}. Further, Trotterization
is not uniquely defined, and we discuss different schemes in Methods. To
establish the formal equivalence with a periodically kicked system, in
Eq.~\eqref{eq:Suzuki-Trotter} we identified the Trotter step size with the
kicking period, $\tau = t/n$. One of the main goals of DQS is to enable studying
the dynamics of quantum many-body systems in regimes where the growth of
entanglement prohibits classical simulations. A typical model system which is
implemented in trapped-ion quantum simulators~\cite{Lanyon57, Barreiro2011,
  Martinez2016, Zhang2017, Britton2012,
  Garttner2017}
is the Ising chain described by
\begin{equation}
  \label{eq:long-range-spin}
  H_x = h_x \sum_{i = 1}^N S_i^x, \quad H_z = J_{z, \alpha} \sum_{i < j
    = 1}^N \frac{S_i^z S_j^z}{\abs{i - j}^{\alpha}}, 
\end{equation}
with spin operators $S_i^{\mu}$ where $\mu = x,y,z$ and $i = 1, \dotsc, N$, and
the power-law coupling strength $J_{z, \alpha}$. For such long-range interacting
systems, extensivity of the Hamiltonian is ensured by the Kac
prescription~\cite{Kac1963},
\begin{equation}
  \label{eq:Kac}
  J_{z, \alpha} = J_z \bigg/ \bigg( \frac{1}{N - 1} \sum_{i < j = 1}^N
  \frac{1}{\abs{i - j}^{\alpha}} \bigg).
\end{equation}
In the limit $\alpha\to 0$, this model describes all-to-all interactions between
spins and becomes equivalent to the quantum kicked top. In the opposite limit of
nearest-neighbor interactions, $\alpha \to \infty$, and with the addition of a
longitudinal field in Eq.~\eqref{eq:long-range-spin}, some of us addressed in
Ref.~\cite{Heyl2019} the crucial question of ``Trotter errors,'' i.e., errors in
DQS due to the approximate factorization
$\e^{-\i H \tau} \approx \e^{-\i H_z \tau} \e^{-\i H_x \tau}$. By performing extensive
numerical simulations, Ref.~\cite{Heyl2019} found that the Trotter error in
observables such as the magnetization exhibits threshold behavior as the Trotter
step size $\tau$ is changed. Below the threshold, the Trotter error admits a
perturbative expansion in $\tau$, while it is uncontrolled above the
threshold. Reference~\cite{Heyl2019} attributed this behavior to a transition
from dynamical localization to a regime showing features of many-body quantum
chaos.

Studies of many-body quantum
chaos~\cite{DAlessio2016} 
typically have to resort to numerics (see
Refs.~\cite{Kos2018, Bertini2018} for exceptions). In single-body quantum chaos,
on the other hand, sophisticated techniques have been developed to establish
deep connections to RMT even analytically. To harness this knowledge, we first
focus here on the all-to-all interacting limit $\alpha \to 0$ of the Hamiltonian
Eq.~\eqref{eq:long-range-spin}. In this limit, the collective spin
$S^2 = S_x^2 + S_y^2 + S_z^2$ with components
$S_{\mu} = \sum_{i = 1}^N S_i^{\mu}$ where $\mu = x,y,z$ becomes a constant of
motion. Accordingly, the $2^N$-dimensional Hilbert space of a system of $N$
spin-\nicefrac{1}{2} can be decomposed into decoupled subspaces of fixed total
spin $S$. In each of these subspaces, the Trotterized
dynamics~\eqref{eq:Suzuki-Trotter} reproduces the dynamics of a kicked top of
size $S$. We focus specifically on the subspace with maximal spin $S = N/2$, in
which the Hamiltonians given in Eq.~\eqref{eq:long-range-spin} reduce to those
in Eq.~\eqref{eq:H-kicked-top} (up to an inconsequential additive shift of the
total energy and a rescaling $J_z \to 2 \left( N + 1 \right) J_z/N$ to
accommodate usual conventions). The dimension $\mathcal{D} = 2 S + 1 = N + 1$ of
this subspace scales linearly with the number of spins $N$. Chaotic motion of
the kicked top is ergodic within this subspace, and it does not explore the full
many-body Hilbert space of dimension $2^N$. In this regard, the kicked top does
not display \emph{many-body} quantum chaos.

We show that in this setting nevertheless observables exhibit the same threshold
behavior as described in Ref.~\cite{Heyl2019} for a generic many-body
system. Moreover, we show that the threshold is determined by the onset of
quantum chaos in the kicked top. While this implies that physical observables in
DQS can remain close to their ``ideal'' values, we find that the fidelity of the
simulated many-body quantum states drops sharply with time for any non-zero
Trotter step in large systems. In view of the variety of possible experimental
implementations of the kicked top listed above, we determine requirements on
system sizes and decoherence rates to observe the threshold behavior. It is an
interesting conceptual question whether in an ideal DQS Trotter errors could be
controlled up to arbitrarily long times and in the thermodynamic limit in which
$N$ and hence $S$ tend to infinity. For collective spin models, we find that
this is indeed the case.

Finally, we connect the results for collective spin systems to the case of
generic many-body systems~\cite{Heyl2019}.  In particular, we explore the
influence of deformations of the kicked top to $\alpha > 0$, where the problem
can no longer be solved on the basis of collective spin operators. Instead, we
resort to exact diagonalization of systems up to $N = 14$, where we recover the
essential features of the kicked top. The threshold in Trotter errors persists
over the entire studied range $0 \leq \alpha \leq 3$ from all-to-all to dipolar
interactions.

\begin{figure*}
  \centering
  \includegraphics[width=\linewidth]{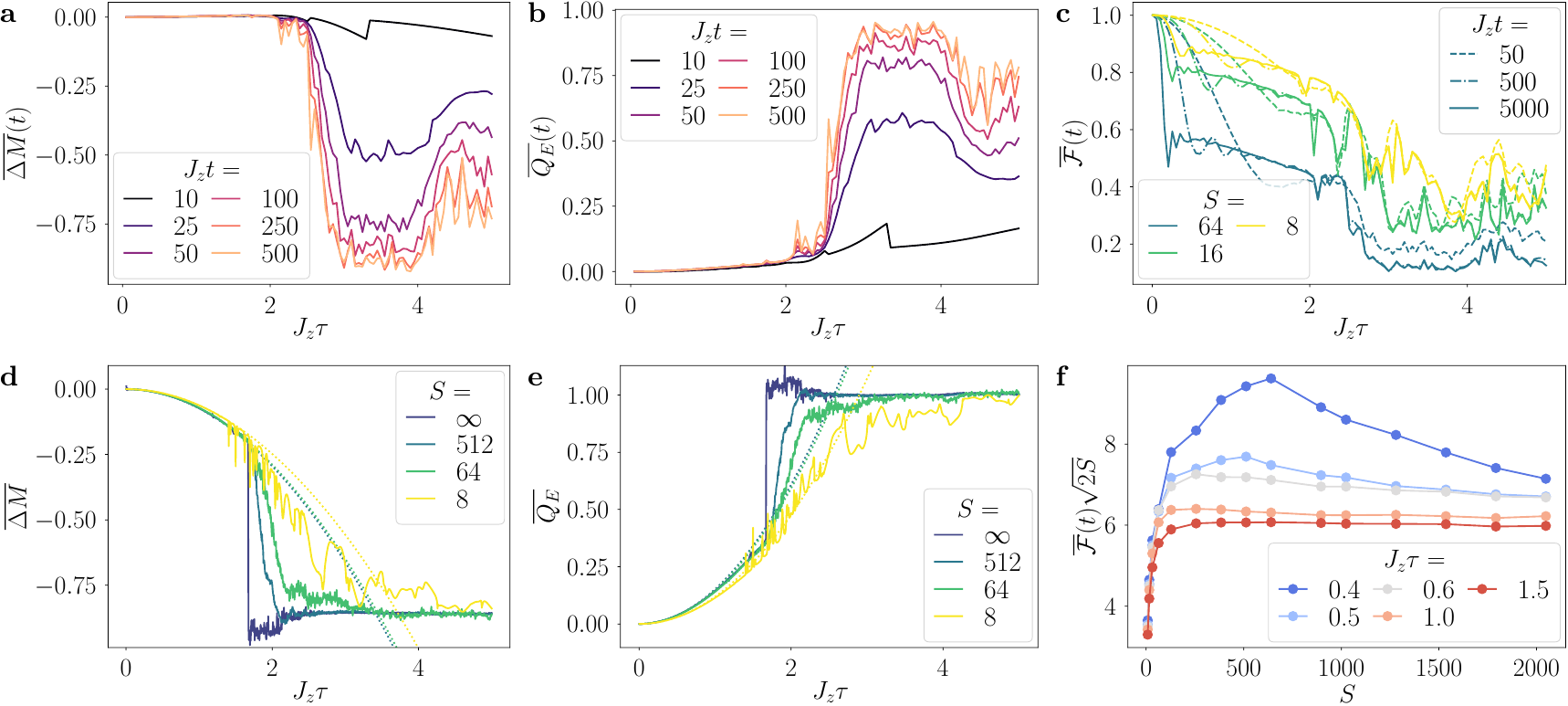}
  \caption{\textbf{a,b} Dynamical buildup of the Trotter-error threshold. For
    the chosen parameters, threshold behavior in \textbf{a} the magnetization
    error $\overline{\Delta M}(t)$ and \textbf{b} heating as quantified by
    $\overline{Q_E}(t)$ sets in at $J_z t_{\mathrm{th}} \approx 25$. The
    discontinuities in the curves pertaining to short times are due to the
    rounding $\left\lfloor t/\tau \right\rfloor$ to integer numbers of Trotter
    steps in the definition of the temporal averages. We chose $S = 50$
    corresponding to recent experiments with trapped ions~\cite{Britton2012,
      Garttner2017}. Smaller (larger) values of $S$ result in a more rounded
    (sharper) threshold. \textbf{c,f} Absence of a sharp threshold in the
    fidelity. In contrast to physical observables, the \textbf{c} time-averaged
    fidelity $\overline{\mathcal{F}}(t)$ does not exhibit persistent threshold
    behavior as a function of the Trotter step size $\tau$. Instead, \textbf{f}
    the fidelity drops with increasing system size as $\sim 1/\sqrt{S}$ even for
    values of $\tau$ within the regular regime. \textbf{d,e} Long-time averages
    of the \textbf{d} magnetization error $\overline{\Delta M}$ and \textbf{e}
    heating as quantified by $\overline{Q_E}$. Dotted lines represent results
    from perturbation theory as explained in the main text. The data for
    $S \rightarrow \infty$ is obtained from the semiclassical equations of
    motion~\eqref{eq:semiclassical-eoms}. For \textbf{a--c} and \textbf{f}, the
    initial state is $\ket{\theta, \phi} = \ket{0, 0} = \ket{S, S}$, while in
    \textbf{d} and \textbf{e} the initial state is
    $\ket{\theta, \phi} = \ket{0.25 \pi, 0}$.}
  \label{fig:Trotter-errors}
\end{figure*}

\sect{Results}
\subsect{Trotter errors in DQS of collective spin systems} We consider two
measures of Trotter errors: deviations of the expectation values of physical
observables~\cite{Heyl2019} and the fidelity of the quantum
state
obtained in DQS. Interestingly, these
measures exhibit strikingly different behaviors.

We first compare the ``magnetization,'' i.e., the expectation value of $S_z$,
under Trotterized and ideal dynamics starting from a spin coherent state
$\ket{\theta, \phi} = \e^{\i \theta \left( S_x \sin(\phi) - S_y \cos(\phi)
  \right)} \ket{S, S_z = S}$.
Here, $\ket{S, S_z}$ is an eigenstate of the spin $z$-component with eigenvalue
$S_z$. The magnetization error is defined as
\begin{equation}
  \label{eq:magnetization-error}
  \Delta M(t) = \frac{1}{S} \left(\langle S_z(t) \rangle_{\tau} - \langle S_z(t)
  \rangle \right),
\end{equation}
where the expectation value $\langle \cdot \rangle_{\tau}$ is taken under
Trotterized dynamics with step size $\tau$ while $\langle \cdot \rangle$
pertains to time evolution with the target Hamiltonian $H$. The accumulated
effect of Trotter errors is quantified by the stroboscopic temporal average
$\overline{\Delta M}(t) = \frac{1}{n_t} \sum_{n = 1}^{n_t} \Delta M(n \tau)$
where $n_t = \left\lfloor t/\tau \right\rfloor$, which is shown in
Fig.~\ref{fig:Trotter-errors}\textbf{a}. For this figure and throughout the paper, we
assume that
\begin{equation}
  \label{eq:collective-spin}
   H_{\mu} = h_{\mu} S_{\mu} + J_{\mu} S_{\mu}^2/(2 S + 1),
\end{equation}
for both $\mu = x$ and $\mu = z$, with $h_x = 0.1 J_z$, $J_x = 0.7 J_z$, and
$h_z = 0.3 J_z$. However, our results do not depend on the precise values of
these parameters (except for fine-tuned points of higher symmetry such as
$h_x = h_z, J_x = J_z$). As shown in the figure, after a characteristic time of
$J_z t_{\mathrm{th}} \approx 25$ for the spin size $S = 50$ and the above
parameters, the time-averaged magnetization error exhibits clear threshold
behavior: While Trotter errors remain small for Trotter step sizes $\tau$ below
a threshold value $J_z \tau_{*} \approx 2.5$, errors become significant for
larger $\tau$. A concrete experimental realization will also suffer from other
imperfections. However, the regimes of bounded and uncontrolled Trotter errors
can still be distinguished if $t_c > t_{\mathrm{th}}$, where $t_c$ is the
coherence time. This is discussed further in Discussion.

As detailed below, $\tau_{*}$ determines the radius of convergence of the
Floquet-Magnus (FM) expansion for the effective Hamiltonian $H_{\tau}$ defined
by $U_{\tau} = \e^{-\i H_{\tau} \tau}$ (see Ref.~\cite{Bandyopadhyay2015} for the
FM expansion in the context of the kicked top). The FM expansion is guaranteed
to converge always if the ``driving frequency'' $\Omega = 2 \pi/\tau$ exceeds
the time-averaged operator norm of the Hamiltonian, i.e.,
$\Omega \gtrsim \frac{1}{\tau} \int_0^{\tau} d t \norm{H(t)}$~\cite{Casas2001}.
Intuitively, for such high driving frequencies, the system cannot absorb energy
from the drive and thus the energy as measured by the expectation value of the
time-averaged Hamiltonian $H = H_x + H_z$ is conserved. The range of values of
the Trotter step $\tau$ in which this criterion applies vanishes for large spin
sizes $S$ since $\norm{H(t)} \sim S$ so that $\tau \lesssim 1/S$.  However, we
find in DQS of collective spin systems that heating is suppressed up to a
Trotter step size $\tau_{*}$ even for large values of $S$. We quantify this
absence of heating for $\tau < \tau_{*}$ by the ``simulation
accuracy''~\cite{Heyl2019}
\begin{equation}
  \label{eq:QE}
  Q_E(t) = \frac{E_{\tau}(t) - E_0}{E_{T = \infty} - E_0}
\end{equation}
where $E_{\tau}(t) = \langle H(t) \rangle_{\tau}$, and
$E_0 = \braket{\psi_0 | H | \psi_0}$ is the energy of the initial state at
$t = 0$. In $Q_E(t)$, the energy increase is normalized with respect to the
difference between the energy of an infinite-temperature state,
$E_{T = \infty} = \tr(H)/\mathcal{D}$ where $\mathcal{D} = 2 S + 1$ is the
Hilbert space dimension, and $E_0$. A value of $Q_E(t) = 1$ thus indicates
heating to infinite temperature. Figure~\ref{fig:Trotter-errors}\textbf{b} shows the
finite-time average,
$\overline{Q_E}(t) = \frac{1}{n_t} \sum_{n = 1}^{n_t} Q_E(n \tau)$, which
clearly exhibits threshold behavior characterized by the same timescales
$t_{\mathrm{th}}$ and $\tau_{*}$ as the magnetization error.

These results demonstrate that few-body observables are quite robust against
Trotter errors for $\tau < \tau_{*}$. This is in stark contrast to the accuracy
of the full unitary time-evolution operator: The difference $U(t) - U_{\tau}^n$,
which quantifies the error made in the Trotterization in
Eq.~\eqref{eq:Suzuki-Trotter}, grows at least linearly in both the simulation
time $t = n \tau$ and the system size $N$~\cite{Berry2007, Jordan2012, Haah2018,
  Childs2019}. Similarly, the robustness of observables also does not extend to
the quantum state $\ket{\psi_{\tau}(t)} = U_{\tau}^{n_t} \ket{\psi_0}$ obtained
under Trotterized dynamics in DQS. In Fig.~\ref{fig:Trotter-errors}\textbf{c}, we show
stroboscopic temporal averages of the fidelity $\mathcal{F}(t)$, i.e., the
absolute value of the overlap of $\ket{\psi_{\tau}(t)}$ with the ideal target
state $\ket{\psi(t)} = U(t) \ket{\psi_0}$,
\begin{equation}
  \label{eq:fidelity}
  \mathcal{F}(t) = \abs{\braket{\psi_{\tau}(t) | \psi(t)}}.
\end{equation}
As illustrated in the figure, the temporal average $\overline{\mathcal{F}}(t)$
approaches the ideal value of $\overline{\mathcal{F}}(t) = 1$ for $\tau \to 0$,
but drops sharply already for $J_z \tau \ll 1$, and in particular for Trotter
step sizes which are much smaller than the threshold value $\tau_{*}$ identified
above for the magnetization error and the simulation accuracy. At this threshold
value, i.e., for $\tau \approx \tau_{*}$, there is another noticeable drop in
$\overline{\mathcal{F}}(t)$. However, also in the region below $\tau_{*}$ the
fidelity vanishes with increasing system size as shown in
Fig.~\ref{fig:Trotter-errors}\textbf{f}. For large system sizes, the decay of the
fidelity approaches $\mathcal{F}(t) \sim 1/\sqrt{\mathcal{D}}$ set by the
Hilbert space dimension $\mathcal{D} = 2 S + 1$~\cite{Emerson2002}. Thus, there
is no persistent threshold behavior in the fidelity of the simulated quantum
state.  In light of this fragility of the quantum state, the robustness of local
observables seems rather remarkable.

Thus far, we have focused on the temporal evolution of Trotter errors and, in
particular, on the dynamical buildup of a threshold in physical
observables. Notably, in collective spin systems, both the infinite-time and
thermodynamic limits are accessible. Figure~\ref{fig:Trotter-errors}\textbf{d} shows
the infinite-time average of the magnetization error. The data for $S < \infty$
is obtained by exact diagonalization of the Floquet operator $U_{\tau}$. In the
limit $S \to \infty$, the effective Planck constant vanishes,
$\hbar_{\mathrm{eff}} = 1/S \to 0$~\cite{Haake2010}. Therefore, in the
thermodynamic limit, the spin components obey semiclassical stroboscopic
evolution equations as detailed in Methods. These evolution equations can be
iterated efficiently, and the long-time average for $S \to \infty$ shown in
Fig.~\ref{fig:Trotter-errors} corresponds to $n = 10^6$ iterations. Fluctuations
in the data for $\overline{\Delta M}$ decrease upon further increasing the
number of iterations.

As can be seen in Fig.~\ref{fig:Trotter-errors}\textbf{d}, the threshold behavior
identified above at finite times persists for $t \to \infty$. We note that the
different shapes of the curves for the magnetization error in
Figs.~\ref{fig:Trotter-errors}\textbf{a} and~\textbf{d} are due to the different choice of the
initial state, which is $\ket{\theta, \phi} = \ket{0, 0} = \ket{S, S}$ and
$\ket{\theta, \phi} = \ket{0.25 \pi, 0}$ in \textbf{a} and \textbf{d}, respectively. We
comment below on these state-dependent variations, also of the threshold value
$\tau_{*}$ itself.

For small Trotter steps, the time-averaged magnetization error admits a
controlled expansion in powers of $\tau$,
\begin{equation}
  \label{eq:Delta-M-bar}
  \overline{\Delta M} = \lim_{n \to \infty} \frac{1}{n} \sum_{n' = 1}^n \Delta
  M(n' \tau) = m_1 \tau + m_2 \tau^2 + O(\tau^3).
\end{equation}
This expansion can be obtained analytically from the FM series for the effective
Hamiltonian by employing time-dependent perturbation theory~\cite{Heyl2019},
which we generalize to arbitrary initial states in the Supplementary
Section 1. While we cannot justify the truncation of the FM expansion underlying
Eq.~\eqref{eq:Delta-M-bar} with full mathematical rigor, we take the excellent
quantitative agreement between Eq.~\eqref{eq:Delta-M-bar} and the numerical data
shown in Fig.~\ref{fig:Trotter-errors}\textbf{d} as evidence that the scaling of the
error predicted by truncating the FM expansion is indeed correct. Our numerical
findings indicate that a fully rigorous justification of the analytical error
estimate can be given, perhaps similarly to the problem of simulating local
lattice Hamiltonians~\cite{Haah2018}, where an integral representation of
Trotter errors for the full time evolution operator~\cite{Childs2019} proved the
correctness of error estimates based on a truncation of the
Baker-Campbell-Hausdorff formula~\cite{Jordan2012}. Further, our findings
suggest that the radius of convergence of the expansion
Eq.~\eqref{eq:Delta-M-bar} coincides with $\tau_{*}$ and is finite.

Figure~\ref{fig:Trotter-errors}\textbf{d} shows data for spin sizes ranging form
$S = 8$ to $S \to \infty$. The smallest value shown, $S = 8$, is realized in
experiments with Dysprosium atoms~~\cite{Chalopin2018}, while $S = 64$ can be
achieved in 1D~\cite{Lanyon57, Zhang2017} or 2D arrays of
ions~\cite{Garttner2017}. A threshold in $\overline{\Delta M}$ is visible over
the entire range of values of $S$. The value of $\tau_{*}$ depends only very
weakly on system size, and crucially it remains finite even for $S \to \infty$.
These observations do not depend qualitatively on the choice of the Hamiltonian
parameters or the angles $\theta$ and $\phi$ specifying the initial coherent
spin state.

In Fig.~\ref{fig:Trotter-errors}\textbf{e}, we also show the long-time average
$\overline{Q_E}$ of the simulation accuracy $Q_E(t)$ defined in
Eq.~\eqref{eq:QE}. Clearly, heating is suppressed to perturbatively small
values, $\overline{Q_E} = q_1 \tau + q_2 \tau^2 + O(\tau^3)$, for Trotter steps
up to the same threshold value $\tau_{*}$ as for the magnetization error,
indicating a finite radius of convergence of the FM expansion even for
$S \to \infty$. As in the expansion for the magnetization error in
Eq.~\eqref{eq:Delta-M-bar}, the coefficients $q_1$ and $q_2$ can be obtained
analytically by time-dependent perturbation theory, see the Supplementary
Section 1.

The existence of controlled perturbative expansions for Trotter errors of
few-body observables up to comparatively large values of the Trotter step size
has profound practical implications for DQS. First, to obtain accurate results
for few-body observables in DQS, one ideally would like to extrapolate from data
at various finite $\tau$ to the limit $\tau \to 0$. Our results show that a
controlled extrapolation is possible from a wide range of values
$\tau < \tau^{*}$ with $J_z \tau^{*}$ of order one, and is not restricted by the
commonly expected requirement $J_z \tau \ll 1$~\cite{Lloyd1073}. Second, the
ability to retain controlled Trotter errors with relatively large Trotter steps
is highly beneficial for current experimental efforts in DQS, as the reduced
number of required quantum gates mitigates the effects of imperfect gate
operations~\cite{Heyl2019}.

The region of controlled Trotter errors at small Trotter step sizes gives way to
a proliferation of errors at $\tau > \tau^{*}$. In particular, the time-averaged
simulation accuracy shown in Fig.~\ref{fig:Trotter-errors}\textbf{e} approaches the
maximum value of $\overline{Q_E} = 1$ with increasing spin size. In Methods we
show that this value is compatible with the assumption that the Floquet
operator~\eqref{eq:U-kicked-top} is a random unitary matrix. In other words, the
Trotterized dynamics becomes completely unrelated to the ideal dynamics
generated by the target Hamiltonian $H$ and in this sense universal. As we
detail in the following, this breakdown of Trotterization, which becomes sharp
for $S \to \infty$, marks the
onset of quantum chaos. \\

\begin{figure*}
  \centering
  \includegraphics[width=\linewidth]{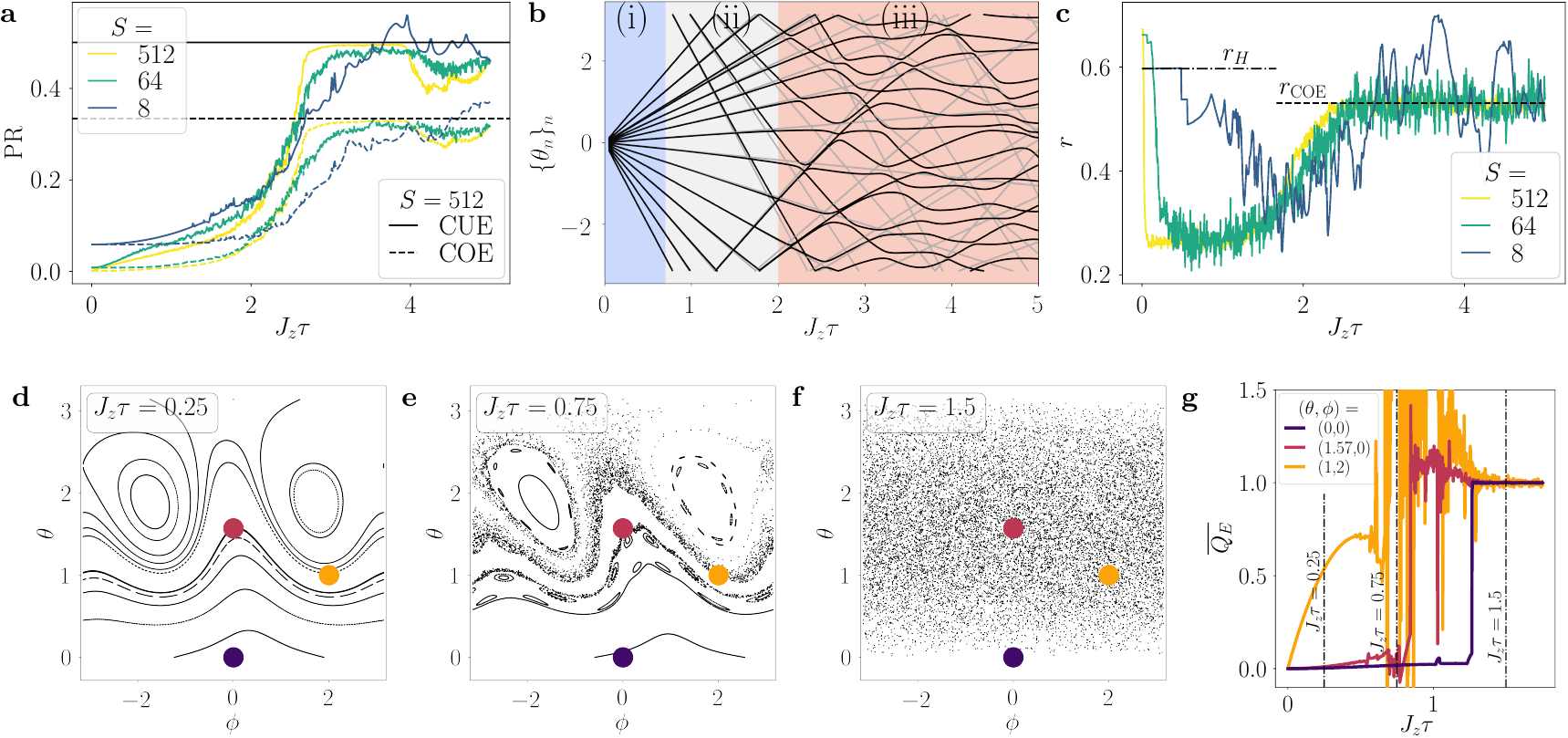}
        \caption{Signatures of quantum chaos: \textbf{a} Participation ratio
          Eq.~\eqref{equ:PR} for the Floquet operator $U_{\tau}$
          Eq.~\eqref{eq:U-kicked-top} (solid lines) and the symmetric Floquet
          operator $U_{\tau, \mathrm{s}}$
          Eq.~\eqref{eq:second-order-Floquet-operator} (dashed
          lines). Horizontal lines pertain to RMT predictions obtained from
          Eqs.~\eqref{eq:CUE-PR} and~\eqref{eq:COE-PR} for a system of size
          $S = 512$.  \textbf{b} Spectrum of the Floquet operator $U_{\tau}$ for
          $S = 8$ with regimes (i), (ii), and (iii) as explained in the main
          text. For comparison, the eigenphases of the ideal time-evolution
          operator $\e^{-\i H \tau}$ are shown as gray lines. \textbf{c}
          Averaged level spacing ratio Eq.~\eqref{equ:AVGphaserepulsion}. The
          horizontal lines in \textbf{c} correspond to COE statistics
          ($r_{\mathrm{COE}}$, dashed), and the value $r_H$ is obtained from the
          time-averaged Hamiltonian $H = H_x + H_z$ for $S=8$
          (dash-dotted). Panels \textbf{a--c} consistently indicate a quantum
          chaos threshold in the dynamics of the quantum top, which lies at the
          core of the proliferation of Trotter errors. \textbf{d--f} Poincar\'e
          sections obtained from the semiclassical equations of
          motion~\eqref{eq:semiclassical-eoms} are shown for Trotter steps
          $J_z \tau=0.25, 0.75, 1.5$, respectively. Depending on the Trotter
          step size, trajectories which emanate from initial states marked by
          the colored dots are regular or chaotic. This is reflected in the
          simulation accuracy $\overline{Q_E}$ shown in \textbf{g}. The simulation
          accuracy is averaged over the first $10^6$ recursion steps.}
  \label{fig:signatures-chaos}
\end{figure*}

\subsect{Interpretation of the proliferation of Trotter errors as quantum chaos}
As we show in the following, the threshold behavior in the magnetization error
and the simulation accuracy can be traced back to the transition from dynamical
localization to quantum chaos in the kicked top. Indeed, for strong kicking,
which corresponds to large Trotter steps $\tau$, the kicked top is well-known to
exhibit quantum chaos~\cite{Haake2010}. Then, statistical properties of the
eigenvectors and eigenvalues of the Floquet operator Eq.~\eqref{eq:U-kicked-top}
are faithful to RMT predictions. In the case of Hamiltonians $H_{x,z}$ as given
in Eq.~\eqref{eq:collective-spin} with all coefficients $h_{x,z}$ and $J_{x,z}$
different from zero, the Floquet operator does not have any symmetries other
than time reversal~\cite{Haake2010}. Thence, the relevant RMT ensemble is the
circular orthogonal ensemble (COE) of unitary and symmetric matrices.

Our first indicator of quantum chaos is connected to the inverse participation
ratio (IPR) for a state $\ket{\psi_0}$,
\begin{equation}
  \label{eq:IPR}
  \mathrm{IPR}(\vert \psi_0 \rangle) = \sum_{m = 1}^{\mathcal{D}} \abs{\langle \psi_0 \vert \phi_m \rangle}^4 
\end{equation}
where $\{\vert \phi_m \rangle \}_m$ denotes the eigenbasis of the Floquet
operator $U_\tau$. We average the IPR over the basis of the target Hamiltonian
$\{\vert \psi_n \rangle \}_n$ and take its inverse to obtain the participation
ratio (PR),
\begin{equation}
  \label{equ:PR}
  \mathrm{PR} = \left( \sum_{n,m = 1}^{\mathcal{D}} \vert \langle \psi_n \vert \phi_m \rangle \vert^4 \right)^{-1}.
\end{equation}
In this definition, we rescaled the PR by a factor of $1/\mathcal{D}$ so that it
assumes values between two limiting cases:
\begin{equation}
  \mathrm{PR} =
  \begin{cases} \mathcal{D}^{-1} & \text{fully localized,} \\
    1 & \text{fully delocalized.}
  \end{cases}
\end{equation}
A low PR thus indicates that the two eigenbases are very similar (localization)
whereas a high PR indicates equal absolute overlaps between all the eigenstates
(delocalization). We show the PR for the kicked top in
Fig.~\ref{fig:signatures-chaos}\textbf{a}. Similar to the Trotter errors discussed
above, the PR exhibits a steep increase at a critical value of the Trotter step
size $\tau$. However, we would like to emphasize a crucial difference between
the PR and Trotter errors of observables: The threshold value $\tau_{*}$ for
Trotter errors depends on the initial state of the time evolution as illustrated
Figs.~\ref{fig:Trotter-errors}\textbf{a,b} and~\textbf{d,e}. On the other hand, the PR as
defined in Eq.~\eqref{equ:PR} contains an average over the entire eigenbasis of
the target Hamiltonian, and consequently it is a ``global'' rather than a
state-dependent and thus ``local'' measure of the onset of chaos. We discuss
this distinction in more detail below.

In Fig.~\ref{fig:signatures-chaos}\textbf{a}, two sets of curves are displayed: Solid
lines pertain to the PR calculated with eigenvectors $\ket{\phi_m}$ of the
Floquet operator $U_{\tau}$ given in Eq.~\eqref{eq:U-kicked-top}; dashed lines
correspond to a symmetric Floquet operator,
\begin{equation}
  \label{eq:second-order-Floquet-operator}
  U_{\tau, \mathrm{s}} = \e^{-\i H_x \tau/2} \e^{-\i H_z \tau} \e^{-\i H_x \tau/2},
\end{equation}
which realizes a higher-order Trotter decomposition as explained in Methods, and
is related to the Floquet operator $U_{\tau}$ in Eq.~\eqref{eq:U-kicked-top} by
a unitary transformation,
\begin{equation}
  \label{eq:first-to-second-order-Trotter}
  U_{\tau} = Q^{\dagger} U_{\tau, \mathrm{s}} Q, \qquad Q = \e^{-\i H_x \tau/2}.
\end{equation}
In the chaotic regime of large Trotter steps, the PR obtained from the Floquet
operator $U_{\tau}$ Eq.~\eqref{eq:U-kicked-top} assumes the value predicted for
the CUE of RMT, see Eq.~\eqref{eq:CUE-PR}. On the other hand, the symmetric
Floquet operator $U_{\tau, \mathrm{s}}$ yields a value of the PR that matches
the COE prediction obtained from Eq.~\eqref{eq:COE-PR}. This difference can be
understood by noting that since the Floquet opreator $U_{\tau, \mathrm{s}}$ is
symmetric, $U_{\tau, \mathrm{s}} = U_{\tau, \mathrm{s}}^T$, it can be regarded
as an element of the COE. The PR Eq.~\eqref{equ:PR} probes the statistics of
eigenvectors of the Floquet operator $U_{\tau, \mathrm{s}}$, and in the chaotic
regime it approaches the RMT prediction. As we explain in Methods, the RMT
prediction is obtained by taking an average over the ensemble of eigenvectors of
random matrices of the COE, i.e., unit-norm vectors with real components.
$U_{\tau, \mathrm{s}}$ is related to the non-symmetric Floquet operator
$U_{\tau}$ by the unitary transformation given in
Eq.~\eqref{eq:first-to-second-order-Trotter}. Since a unitary transformation
does not affect the spectrum of an operator, both $U_{\tau}$ and
$U_{\tau, \mathrm{s}}$ exhibit the same level statistics in the chaotic regime
which we discuss in more detail below. However, as shown in
Fig.~\ref{fig:signatures-chaos}\textbf{a}, the unitary transformation $Q$ strongly
affects the PR. Indeed, our results for the PR corresponding to $U_{\tau}$
indicate that applying $Q$ to the ensemble of eigenvectors corresponding to the
COE yields an ensemble of vectors that resembles the one of the CUE, which is
comprised of unit-norm vectors with complex components.

At small Trotter steps, the PR does not assume the lowest possible value
$\sim 1/\mathcal{D} = 1/(2 S + 1) = 1/(N + 1)$. Instead, it appears to converge
to a finite value, which indicates that even in this localized regime the basis
states $\ket{\psi_m}$ of $H$ are spread out over $\sim \mathcal{D}$ Floquet
states $\ket{\phi_m}$. This is similar to the many-body case considered in
Ref.~\cite{Heyl2019} and further below, where for small $\tau$ an exponential
number of Floquet states is required to represent generic states, but with an
exponent that is smaller than the maximal possible one.

Signatures of the transition to chaos can also be found in the spectrum
$\sigma(U_\tau)$ of the Floquet operator. The eigenphases $\{\theta_n\}_n$ of
the unitary operator $U_\tau$ are defined through
$\sigma(U_\tau) = \{\e^{\i \theta_n} \}_n$ and are shown in
Fig.~\ref{fig:signatures-chaos}\textbf{b} for a system with $S = 8$. Already by visual
inspection of the figure, three regimes can be distinguished: (i) In the region
of very small $\tau$, the eigenphases of the Floquet operator $U_{\tau}$
coincide with the phases corresponding to the ideal time-evolution operator over
one Trotter step, $\e^{-\i H \tau}$, which are shown as gray lines in the
figure. (ii) In the second regime, the phases start to wind around the interval
$[-\pi,\pi]$. This leads to level crossings, which, however, are avoided ones
with only very narrow gaps. (iii) A dramatic change can be observed in the third
regime, where phase repulsion becomes strongly pronounced, which indicates
significant overlaps between the respective eigenstates.

A quantitative one-parameter measure of the degree of phase repulsion is given
by the average adjacent phase spacing ratio $r$~\cite{Oganesyan2007}. In terms
of the phase spacings $\delta_n = \theta_{n+1} - \theta_n$, the average spacing
ratio is defined through
\begin{equation}
  \label{equ:AVGphaserepulsion}
  r_n = \frac{\min(\delta_n,
    \delta_{n+1})}{\max(\delta_n, \delta_{n+1})}, \qquad r =
  \frac{1}{\mathcal{D}} \sum_{n=1}^\mathcal{D} r_n
\end{equation}
where $\mathcal{D} = 2 S + 1$ denotes the dimension of the Hilbert space. The
phase spacing ratio takes universal values for distinct ensembles of random
matrices: In the absence of correlations between the eigenphases, which
corresponds to Poissonian statistics of the adjacent phase spacings, the phase
spacing ratio is given by $r_{\mathrm{POI}} = 2 \ln(2) - 1 \approx 0.39$; In
contrast, the repulsion between eigenphases of random matrices sampled from the
COE is reflected in a higher universal value of the adjacent phase spacing
ratio, $r_{\mathrm{COE}} = 0.5996(1)$~\cite{Atas2013}. For the Floquet operator
$U_{\tau}$ in Eq.~\eqref{eq:U-kicked-top}, the adjacent phase spacing ratio is
shown in Fig.~\ref{fig:signatures-chaos}\textbf{c}. Again, three regimes can be
distinguished: In a region of size $\sim 1/S$ in which the spectral width of the
target Hamiltonian $H$ is smaller than $\Omega = 2 \pi/\tau$, $r$ is determined
by the spectral statistics of $H$. The corresponding value of $r$ depends on the
spin size $S$ and does not assume any of the universal ones given above. This is
not surprising: It is well-known that quantum systems with time-independent
Hamiltonians that have classical counterparts with only one degree of freedom
exhibit non-universal level statistics~\cite{Haake2010, DAlessio2016}. This is
the case, in particular, for the considered spin systems, where the expectation
values of the collective spin operators span a two-dimensional phase space
corresponding to a single degree of freedom in the classical limit. Indeed, for
a classical top $S$ with spin components $S_{x, y, z}$, the dynamics is
restricted to the surface of a sphere, $S^2 = S_x^2 + S_y^2 + S_z^2$, and can be
parameterized in terms of the polar and azimuthal angles $\theta$ and $\phi$,
respectively, which span the two-dimensional phase space
$\left( \theta, \phi \right) \in [0, \pi) \times [0, 2 \pi)$. In the second
regime in Fig.~\ref{fig:signatures-chaos}\textbf{c}, the eigenphases $\theta_n$
of the Floquet operator wind around the unit circle. As shown in the previous
section, for values of $\tau$ up to $J_z \tau \lesssim 2$, the dynamics is
well-captured by the time-independent effective Hamiltonian $H_{\tau}$ obtained
in the FM expansion. For the average phase spacing ratio we find
$r \approx 0.26$ in this regime. This non-universal value depends on the choice
of parameters $h_{x, z}$ and $J_{x, z}$. Finally, for large values of $\tau$,
the adjacent phase spacing ratio assumes the universal value predicted by RMT
for the COE.

We note that while the behavior of few-body observables considered above is
exactly analogous to the phenomenology of generic (short-range interacting)
many-body systems~\cite{Heyl2019}, here we find a crucial difference: Even in
the dynamically localized regime, the time-evolution operator of an interacting
many-body system displays COE phase statistics~\cite{DAlessio2014b}. Therefore,
the statistics of the eigenphases of the Floquet operator does not allow one to
distinguish between localized and many-body chaotic regimes.

The participation ratio and the adjacent gap ratio include averages over the
entire set of eigenstates and eigenphases of the Floquet operator $U_{\tau}$ in
Eq.~\eqref{eq:U-kicked-top}. Hence, they pertain to ``global'' properties of the
dynamics, and show that at large Trotter step sizes the Floquet operator is
faithful to RMT---a hallmark of single-body quantum chaos. In contrast, the
magnetization error and the simulation accuracy discussed above measure Trotter
errors in the time evolution starting from a single initial state. Therefore,
they yield a more ``local'' measure of irregularity in the dynamics, and indeed
we found that the value $\tau_{*}$ of the threshold in Trotter errors depends on
the initial state. To explain this initial-state dependence, we consider again
the semiclassical limit of the top. From the stroboscopic Heisenberg equations
for the spin-components of the top, we obtain in the semiclassical limit the
Poincar\'e sections shown in Fig.~\ref{fig:signatures-chaos}\textbf{d--f}. Regimes of
regular and chaotic dynamics coexist in the classical phase space, and their
relative weight is tuned by the Trotter step size $\tau$. This coexistence is
reflected in the initial-state dependence of $\tau_*$ in the thermodynamic limit
as can be seen in Fig.~\ref{fig:signatures-chaos}\textbf{g}, where we show the
simulation accuracy for trajectories which emanate from the initial states
marked by colored dots in Fig.~\ref{fig:signatures-chaos}\textbf{d--f}. Note that for
our choice of parameters $h_{x, z}$ and $J_{x, z}$, states pertaining to
$\theta \in \{0, \pi\}$ are particularly stable whereas states around
$\theta \approx \frac{\pi}{2}$ are generally more sensitive.

The signatures of chaos discussed so far show clearly that the proliferation of
Trotter errors is a concomitant effect of the transition from regular to chaotic
dynamics in the kicked top. However, in contrast to Trotter errors, in
experiments these signatures are not directly accessible. We proceed to discuss
how chaos is quantified by out-of-time-ordered correlation functions (OTOCs)
which, as we show in the Supplementary Section 2,
can be measured in experimental realizations of the kicked top. \\

\begin{figure*}
  \centering
  \includegraphics[width=\linewidth]{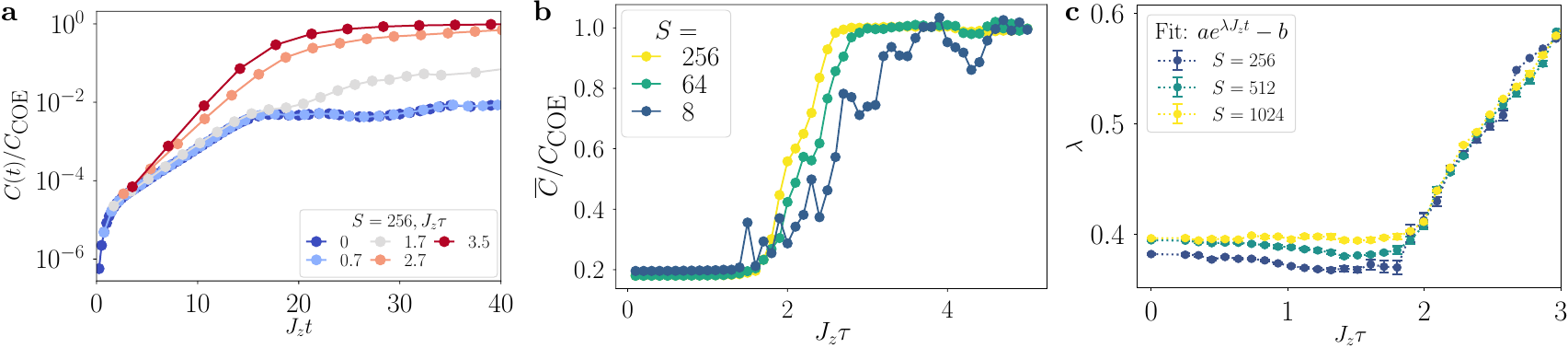}
        \caption{Signatures of quantum chaos in OTOCs. \textbf{a} Short-time growth of
          the infinite-temperature OTOC Eq.~\eqref{eq:otoc} and \textbf{b}
          infinite-time average $\overline{C}$. In the chaotic regime, the
          exponential growth of $C(t)$ saturates to the COE value obtained from
          Eq.~\eqref{eq:otoc-coe}. The infinite-time average $\overline{C}$
          follows from Eqs.~\eqref{eq:C-F-otoc}
          and~\eqref{eq:F-otoc-infinite-time-avg}. \textbf{c} Growth rate $\lambda$ for
          the range of times in \textbf{a} during which the OTOC grows
          exponentially. The rate $\lambda$ is extracted from the data by a fit
          to $C(t) = a \e^{\lambda J_z t} - b,$ where $a$, $b$, and $\lambda$ are
          fit parameters. Error bars correspond to fitting error. In the regular
          regime, the growth rate takes a value that is independent of the
          Trotter step size $\tau$ and agrees with the value at $J_z \tau = 0$
          which we obtained from the ideal evolution with the time-averaged
          Hamiltonian. For Trotter step sizes in the chaotic region, the growth
          rate increases with $\tau$.}
  \label{fig:otocs}
\end{figure*}

\subsect{Signatures of dynamical localization and quantum chaos in
  out-of-time-ordered correlators} A hallmark of chaotic dynamics in classical
systems is the butterfly effect, which is caused by extreme sensitivity to
initial conditions: In a chaotic system, two trajectories that emanate from
nearby points in phase space deviate from each other exponentially fast. The
rate entering this exponential, which is called the Lyapunov exponent, is a
quantitative measure of chaos.

A generalization of these concepts to the quantum domain can be given in terms
of OTOCs~\cite{Larkin1969, Maldacena2016a, Kitaev}. The ``infinite-temperature''
OTOC for two initially commuting operators $V$ and $W$ is defined as
\begin{equation}
  \label{eq:otoc}  
  C(t) = \frac{1}{\mathcal{D}} \tr \! \left( \abs{[W(t), V]}^2 \right)
  ,  
\end{equation}
where
$\mathcal{D}$ is the Hilbert space dimension, and the Heisenberg operator
$W(t) = U(t)^{\dagger} W U(t)$ is evolved unitarily with $U(t)$. Here, we
consider a single spin $S$ with Trotterized dynamics, for which the Hilbert
space dimension is $\mathcal{D} = 2 S + 1$ and we evaluate $C(t)$
stroboscopically at multiples of the Trotter step such that $t = n \tau$ and
$W(n \tau) = U_{\tau}^{n \dagger} W U_{\tau}^n$. In certain highly chaotic
systems~\cite{Maldacena2016a, Kitaev, Hosur2016}, the OTOC exhibits an extended
period of exponential growth, and the growth rate serves as a quantum analog of
the classical Lyapunov exponent. Moreover, the infinite-time limiting value of
the OTOC is maximal if the time-evolution operator $U(t)$ is faithful to an
ensemble of RMT~\cite{Roberts2017}. Due to its properties as a diagnostic of
chaos, the OTOC has gained a lot of attention recently in a wide range of
physical contexts including high-energy, condensed matter, and AMO
physics~\cite{Maldacena2016a, Kitaev, Hosur2016, Pappalardi2018, Rozenbaum2017a,
  Swingle2016, Garttner2017}
. In experimental
realizations of the kicked top, the OTOC could be accessed through a recently
introduced scheme which is based on correlations between randomized
measurements~\cite{Vermersch2018a} and has already been demonstrated in
NMR~\cite{Nie2019}. We give details on the experimental requirements in the
Supplementary Section 2.

As we show in the following, in the kicked top, both the short-time growth as
well as the long-time saturation of the OTOC are indicative of the transition
from dynamical localization to chaos. This is illustrated in
Fig.~\ref{fig:otocs}, which shows the infinite-temperature OTOC~\eqref{eq:otoc}
with unitary $V = \e^{-\i \varphi S_z}$ and Hermitian $W = S_z$---a particular
choice of operators that facilitates the measurement of
$C(t)$~\cite{Swingle2016, Zhu2016, Shen2017, Yoshida2018, Yao2016, Li2017,
  Garttner2017, Vermersch2018a} as also discussed in the Supplementary
Section 2. However, we note that for the value $\varphi = 10^{-4}$ shown in
Fig.~\ref{fig:otocs}, the OTOC reduces to the form
$C(t) = - \frac{\varphi^2}{\mathcal{D}} \tr ( [S_z(t), S_z]^2 ) + O(\varphi^4)$.

The growth of the OTOC is depicted in Fig.~\ref{fig:otocs}\textbf{a}. As expected in a
chaotic system, for large Trotter step sizes the OTOC grows exponentially,
$C(t) \sim \e^{\lambda J_z t}$~\cite{Pappalardi2018, Seshadri2018}. This
exponential growth extends up to the Ehrenfest time
$t_{\mathrm{E}} \sim \abs{\ln(\hbar_{\mathrm{eff}})}$,
where the effective Planck constant is determined by the spin size,
$\hbar_{\mathrm{eff}} = 1/S$~\cite{Haake2010}. For times $t$ beyond the
Ehrenfest time $t_{\mathrm{E}}$, the OTOC saturates at a value
$C(t) = C_{\mathrm{COE}}$ that is compatible with replacing the time-evolution
operator in Eq.~\eqref{eq:otoc} by a random unitary matrix drawn from the COE as
shown in Methods.

The behavior of the OTOC is markedly different in the regular regime. In
particular, for the smallest non-zero Trotter step size $J_z \tau = 0.73$ shown
in Fig.~\ref{fig:otocs}\textbf{a}, the OTOC is essentially indistinguishable from
evolution under $U(t) = \e^{-\i H t}$ with the target Hamiltonian $H = H_x + H_z$,
which is labeled as $J_z \tau = 0$ in the figure. From a short-time expansion of
the time evolution operator as $\e^{-\i H t} = 1 - \i H t + O(t^2)$ it follows
immediately that at very short times the OTOC grows algebraically,
$C(t) \sim t^2$. The algebraic behavior persists approximately up to a time
$J_z t$ of order one.
Then, the OTOC grows exponentially also in the regular regime. Similar
behavior, i.e., exponential growth of the OTOC in a (classically) regular
parameter regime, was recently discussed in Ref.~\cite{Rozenbaum2017a} for the
kicked rotor. However, there are two key differences between the regular and
chaotic regimes: (i) In the regular regime, the exponential growth gives way to
much slower growth at late times, and eventually saturates at a value below
$C_{\mathrm{COE}}$ as shown in Fig.~\ref{fig:otocs}\textbf{b}; (ii) Moreover, the
growth rate $\lambda$, which we extract from the numerical data in the region of
exponential growth, assumes a constant but weakly system-size dependent value in
the regular regime as shown in Fig.~\ref{fig:otocs}\textbf{c}. In contrast, it rises
sharply upon increasing the Trotter step size $\tau$ beyond the threshold to
chaotic dynamics.

In Fig.~\ref{fig:otocs}, we show data for spin sizes $S = 256, 512, 1024$. These
rather large system sizes allow us to determine the growth rate $\lambda$ very
accurately due to the large Ehrenfest time $t_{\mathrm{E}} \sim \ln(S)$.
However, we found (see Supplementary Section 2) that the two regimes in the
dependence of $\lambda$ on the Trotter step
size---$\lambda \approx \mathrm{const.}$ in the regular regime and strong
increase of $\lambda$ with $\tau$ in the chaotic regime---can be clearly
distinguished also in systems with $S \gtrsim 50$ as in recent measurements of
OTOCs with trapped ions~\cite{Garttner2017}. \\

\begin{figure*}
  \centering
  \includegraphics[width=\linewidth]{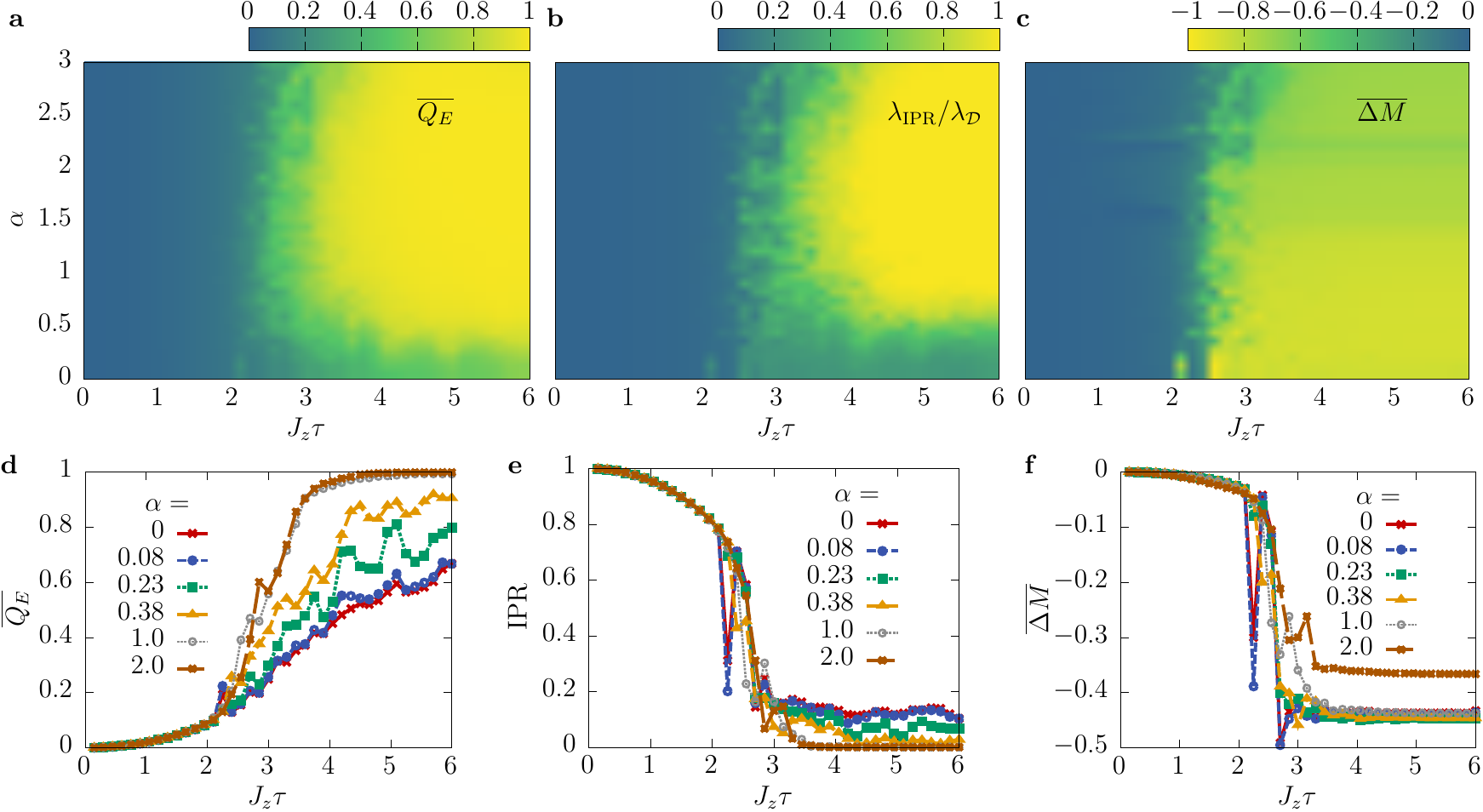}
  \caption{Many-body quantum chaos threshold in a kicked long-range Ising model
    with algebraically decaying interactions set by the interaction exponent
    $\alpha$ for a spin chain of $N = 14$ lattice sites. \textbf{a} Long-time average
    of the simulation accuracy $\overline{Q_E}$ as a function of the Trotter
    step size $\tau$ and $\alpha$. \textbf{b} Rate function $\lambda_\mathrm{IPR}$ of
    the inverse-participation ratio normalized with respect to the maximally
    reachable value $\lambda_\mathcal{D}$ in a fully quantum chaotic case. \textbf{c}
    Long-time limit of the time-averaged magnetization error
    $\overline{\Delta M}$. Although finite-size corrections are significant for
    $0<\alpha\lesssim 1/4$, a clear quantum-chaos threshold can be discerned
    that coincides with the proliferation of Trotter errors on local
    observables. Panels \textbf{d--f} show cuts for the data presented in \textbf{a--c} for
    fixed values of $\alpha$. The quantities shown illustrate clearly the smooth
    crossover from many-body quantum chaos at finite $\alpha$ to the single-body
    chaos of the kicked top at $\alpha = 0$.  In \textbf{e}, a sharp threshold is
    visible in the IPR even at small $\alpha$. For the larger values of $\alpha$
    shown in \textbf{b} the rate function $\lambda_{\mathrm{IPR}}$ is the appropriate
    measure of localization vs.~delocalization over an exponentially large
    Hilbert space.}
  \label{fig:lrising_quantumchaos}
\end{figure*}

\subsect{The kicked top as a limit of long-range interacting spin chains} In the
previous sections, we discussed extensively the kicked top as an ideal model
system for the study of fundamental questions of both DQS and the associated
quantum chaos threshold. Now we explore to which extent our observations persist
under natural deformations of the Hamiltonian in experimentally relevant
settings. For this purpose, we consider Ising chains with algebraically decaying
instead of all-to-all interactions, which exhibit a direct realization in
trapped-ion quantum computers~\cite{Lanyon57, Barreiro2011,
  Martinez2016, Zhang2017, Britton2012,
  Garttner2017}
. Specifically, we study a kicked spin chain as
defined in Eqs.~\eqref{eq:long-range-spin} and~\eqref{eq:Kac}. By tuning the
exponent $\alpha$, the long-range Ising model allows us to smoothly interpolate
between the two paradigmatic model systems of the infinite-range kicked top
studied above and the nearest-neighbor Ising chain, reached at $\alpha \to 0$
and $\alpha\to\infty$, respectively. Motivated by realistic trapped-ion
experiments, we consider open boundary conditions. For the numerical data shown
in what follows, we choose $h_x/J_z = 1/4$, but the main picture does not change
for different parameter sets. We use a Lanczos algorithm with full
reorthogonalization to propagate the initial state
$\ket{\psi_0} = \bigotimes_{i = 1}^N \ket{\uparrow}$ over up to $2 \cdot 10^4$
periods. For the quantities shown below we perform a temporal average over all
periods.

As explained above, the many-body Hilbert space of $N$ spin-\nicefrac{1}{2} can
be decomposed into subspaces with fixed total spin $S$. The initial state
$\ket{\psi_0}$ given above belongs to the subspace of maximal spin $S = N/2$,
and for $\alpha \to 0$ the dynamics is restricted to this subspace. That is, for
$\alpha \to 0$ the Trotterized dynamics of the spin chain as defined in
Eqs.~\eqref{eq:long-range-spin} and~\eqref{eq:Kac} becomes equivalent to the
kicked top Eq.~\eqref{eq:H-kicked-top} with $S = N/2$ and a rescaling of the
parameter $J_z$ as pointed out above. On the other hand, for non-zero values of
$\alpha$, the dynamics explores the $2^N$-dimensional many-body Hilbert space.

In Fig.~\ref{fig:lrising_quantumchaos}\textbf{a}, we plot the numerically obtained data
for the simulation accuracy $Q_E$ defined in Eq.~\eqref{eq:QE} in the long-time
limit as a function of Trotter step $\tau$ and interaction exponent $\alpha$ for
a fixed system size $N=14$. To clearly illustrate how the kicked top is
approached in the limit $\alpha \to 0$, Fig.~\ref{fig:lrising_quantumchaos}\textbf{d}
shows cuts through the same data for four values of the exponent $\alpha$ form
$\alpha = 0$ to $\alpha = 0.38$. For small $\tau$, we find that the properties
observed already with the kicked top extend to $\alpha>0$. Specifically,
$\overline{Q_E}$ is small with $\overline{Q_E} = q_2 \tau^2 + O(\tau^3)$ (the
first-order term vanishes for the initial state
$\ket{\psi_0} = \bigotimes_{i = 1}^N \ket{\uparrow}$). Upon increasing the
Trotter step for $\alpha \lesssim 1/4$, however, $\overline{Q_E}$ does not reach
the value $\overline{Q_E} \to 1$ expected in the quantum chaotic regime. We
attribute this to the small number $N=14$ of simulated spins and consequently to
large finite-size corrections. This agrees also with our observations in
Fig.~\ref{fig:Trotter-errors}\textbf{e} for the kicked top at $\alpha=0$, where
significant finite-size effects in $\overline{Q_E}$ appear (note also the
different choice of initial state and Hamiltonian parameters in
Fig.~\ref{fig:Trotter-errors}\textbf{e}). Instead, for $\alpha \gtrsim 1/4$, we find a
clear crossover from a perturbative regime at small Trotter steps $\tau$,
implying controllable Trotter errors, to a many-body quantum chaotic region with
$\overline{Q_E} \to 1$. Except for the range $0 \leq \alpha \lesssim 1/4$, the
influence of the interaction exponent $\alpha$ on the simulation accuracy is
only minor in that also the crossover region only shifts slightly upon varying
$\alpha$.

A similar picture emerges from the IPR defined in Eq.~\eqref{eq:IPR}, which
measures the localization properties of the state $\ket{\psi_0}$ in the
eigenbasis $\{ \ket{\phi_m} \}_m$ of the Floquet operator. For $\alpha > 0$, the
accessible Hilbert space grows exponentially with the number of spins,
$\mathcal{D} = 2^N$. To account for this exponential dependence, we introduce
the rate function
$\lambda_{\mathrm{IPR}} = \ln(\mathrm{IPR})/N$~\cite{Heyl2019}. In
Fig.~\ref{fig:lrising_quantumchaos}\textbf{b}, we show the ratio
$\lambda_{\mathrm{IPR}}/\lambda_{\mathcal{D}}$, where
$\lambda_{\mathcal{D}} = \ln(2) \left( 1 - 3/N \right)$. This value incorporates
leading-order finite-size corrections as follows: On the one hand the system
obeys both an inversion as well as Ising $\mathbb{Z}_2$ symmetry, such that
$\mathcal{D}=2^{N-2}$; On the other hand, RMT allows us to estimate the IPR in
an ergodic phase as $\mathrm{IPR} \sim 2/\mathcal{D}$ for
$\mathcal{D} \to \infty$, see Methods. Combining these two results we obtain the
value of $\lambda_\mathcal{D}$ given above. We note that here we restrict
ourselves to comparing the IPR obtained for the initial state $\ket{\psi_0}$ to
the RMT prediction for the CUE Eq.~\eqref{eq:CUE-PR}. As discussed above, we
expect that a symmetrized Floquet operator yields an IPR that is consistent with
the COE result~\eqref{eq:COE-PR}.

The rate function $\lambda_{\mathrm{IPR}}$ essentially reproduces the
observations from the simulation accuracy in the range $0<\alpha \leq 3$ with
only slight modifications, which we again attribute to finite-size
corrections. The case $\alpha$ close to $\alpha=0$, however, deserves a more
detailed discussion. While a system at $\alpha=0$ with total spin $S = N/2$ can
access only $\mathcal{D} = 2S+1 = N + 1$ quantum states in the Hilbert space,
with any infinitesimal deviation from $\alpha=0$ this number grows to
$\mathcal{D} = 2^N$. For a system with finite $N$ as we consider here, however,
we expect that the effective behavior of $\mathcal{D}$ has to interpolate from
the polynomial dependence at $\alpha=0$ to the exponential dependence in $N$
upon increasing $\alpha$. This opens up a crossover window at small $\alpha$,
whose size diminishes for larger $N$ and where the rate function
$\lambda_{\mathrm{IPR}}$ exhibits strong finite-size corrections, as is also
visible in Fig.~\ref{fig:lrising_quantumchaos}\textbf{b}. Within the crossover window,
the threshold to chaos is nevertheless clearly visible if instead of the rate
function $\lambda_{\mathrm{IPR}}$ one considers the IPR Eq.~\eqref{eq:IPR}
itself. This is illustrated in Fig.~\ref{fig:lrising_quantumchaos}\textbf{e}, which is
based on the same data as panel \textbf{b} and shows cuts for several small values of
$\alpha$. Beyond the crossover window, the rate function shown in
Fig.~\ref{fig:lrising_quantumchaos}\textbf{b} reproduces the observations from the
simulation accuracy, while the crossover region from quantum localized to
chaotic appears slightly broader.

The previous considerations provide numerical evidence for a quantum many-body
chaos transition in the Trotterized time evolution. For completeness, we now aim
to confirm our general arguments on controllable Trotter errors by studying also
the long-time limit of a local observable, namely the magnetization $M$ as done
also for the kicked top. Our numerical data for the time-averaged magnetization
error Eq.~\eqref{eq:magnetization-error} is shown in
Fig.~\ref{fig:lrising_quantumchaos}\textbf{c}. Again, we observe two different regimes
depending on the Trotter step $\tau$. Compared to the simulation accuracy and
the IPR in Fig.~\ref{fig:lrising_quantumchaos}\textbf{b}, the crossover region appears
much sharper, which is also reflected in the cuts through the data shown in
Fig.~\ref{fig:lrising_quantumchaos}\textbf{f}. The overall behavior remains similar:
Delocalization of the time-evolved state in Hilbert space implies an error
$\overline{\Delta M}$ of order one as we find across a wide region in the
$\alpha$-$\tau$ plane. For small values of $\tau$ instead, $\overline{\Delta M}$
is perturbatively small and Trotter errors remain controllable.

Overall, this analysis suggests that a continuous deformation of the kicked top
Hamiltonian to algebraically decaying interactions does not influence our
general observations in the previous parts of the manuscript, while only details
such as the location of the quantum many-body chaos threshold can exhibit slight
modifications. The present study of the regime $0 \leq \alpha \leq 3$ in
combination with the results of Ref.~\cite{Heyl2019}, which considered a
short-range interacting system corresponding to the limit $\alpha \to \infty$,
indicate that the transition to chaos is a generic phenomenon with broad
relevance for DQS.

\sect{Discussion} In the field of quantum chaos, it is well-known that the
kicked top exhibits a transition from regular to chaotic dynamics upon
increasing the kicking strength, and the signatures of this transition, e.g., in
the spectral statistics of the Floquet operator, are well-understood. Our work
connects these results to and highlights their fundamental importance for the
seemingly remote field of quantum information science. Namely, we show that the
transition to chaos becomes manifest in sharp threshold behavior of Trotter
errors of few-body observables in DQS of collective spin systems. Contrary to
the difference between the ideal and the Trotterized time-evolution operator,
which is often used as a measure for Trotter errors in DQS and which grows at
best linearly with simulation time~\cite{Berry2007, Jordan2012, Haah2018,
  Childs2019}, Trotter errors of observables remain controlled and constant up
to arbitrarily long times. A quantitatively accurate description of these
Trotter errors is provided in the regular regime by the FM expansion---even up
to Trotter step sizes, for which one would expect the FM expansion to
diverge~\cite{Casas2001}. In the chaotic regime, the kicked top is well-known to
be faithful to RMT~\cite{Haake2010}. 

It is worthwhile to emphasize again the distinction between single-particle and
many-body quantum chaos, which is central to the results obtained in this
work. On one hand, the quantum kicked top has emerged as a paradigmatic model
for single-particle quantum chaos upon increasing the kicking strength. In the
semiclassical limit, the problem becomes chaotic in the classical sense of
exponentially diverging trajectories upon slightly perturbing initial
conditions. From a quantum many-body perspective, however, the kicked top is an
integrable and therefore non-ergodic system, because its dynamics is restricted
to a $N + 1$-dimensional subspace of the $2^N$-dimensional many-body Hilbert
space. In this regard, the kicked top is also a rather peculiar system for DQS,
which aims at simulating quantum many-body chaotic dynamics in regimes in which
the growth of entanglement prohibits efficient classical simulations.
Nevertheless, in this work we show that the phenomenology of Trotter errors in
this particular DQS is analogous to the behavior found in a generic many-body
system in Ref.~\cite{Heyl2019}, can be explained in terms of quantum chaos in
the kicked top, and is smoothly connected to ``true'' quantum many-body systems
described by Eq.~\eqref{eq:long-range-spin} with $\alpha > 0$.

We note that for small $\alpha > 0$ it is not yet fully understoood whether the
long-range deformation in Eq.~\eqref{eq:long-range-spin} becomes ergodic. In the
kicked top, i.e., for $\alpha = 0$, threshold behavior persists even in the
thermodynamic limit and up to arbitrarily long times. On the other hand, in
generic many-body systems such as the Ising chain Eq.~\eqref{eq:long-range-spin}
with sufficiently large $\alpha$, recent works argued that in the thermodynamic
limit periodic driving will always lead to indefinite heating with the
simulation accuracy Eq.~\eqref{eq:QE} $Q_E \to 1$~\cite{DAlessio2014b,
  Lazarides2014, Luitz2017}, while heating can be strongly suppressed with
$Q_E < 1$ during a prethermal regime~\cite{Abanin2015b, Mori2016, Kuwahara2016,
  Machado2017, Howell2018}. This regime
persists up to a time scale which grows exponentially with the driving
frequency, and has recently been observed experimentally~\cite{Singh2018}. It is
an interesting question for future studies how the above scenarios are connected
upon tuning $\alpha$. Irrespective of this question, the fact that the heating
time scale is infinite in the kicked top and can be made arbitrarily large in
generic short-range interacting systems implies that DQS experiments are not
limited by Trotter errors but rather by extrinsic error sources such as qubit
decoherence. We emphasize that this conclusion is not restricted to all-to-all
interacting systems. Rather, we expect it to hold generically for a large class
of model systems which are relevant for DQS.

Our theoretical predictions concerning Trotter errors and the onset of quantum
chaos can be tested experimentally in a variety of different systems ranging
from single atomic spins~\cite{Chalopin2018, Baier2018} to quantum simulators of
interacting spin-\nicefrac{1}{2} systems~\cite{Lanyon57, Peng2005, Barends2016,
  Langford2016, OMalley2016, Zhang2017, Barends2015, Barreiro2011, Martinez2016,
  Salathe2015}
. A first experimental requirement is the
ability to engineer non-linear spin Hamiltonians with time-dependent
coefficients. Moreover, in any experimental realization, decoherence will
eventually wash out the threshold shown in Fig.~\ref{fig:Trotter-errors}:
Experimental imperfections such as technical noise or coupling of the system to
its environment will lead to further heating and drive the system towards a
featureless infinite-temperature state. In such a state, expectation values of
observables become independent of the Trotter step size and the sharp threshold
is destroyed. Therefore, the decoherence time $t_c$ should be longer than the
time scale $t_{\mathrm{th}}$ it takes to build up the threshold. For the
parameters chosen in Fig.~\ref{fig:Trotter-errors}, this requirement is
$J_z t_{\mathrm{th}} \approx 25 < J_z t_c$ and can be met in current
experiments. (To name an example, Ref.~\cite{Garttner2017} achieved $J_z$ on the
order of several kHz and thus nearly two orders of magnitude larger than the
decoherence rate $1/t_c$ of several ten Hz.) It is an intriguing prospect to
experimentally connect the threshold in Trotter errors to quantum chaos by
measuring the growth rate and saturation value of the OTOC.

To summarize, in this work, we connect the well-studied quantum chaos of the
kicked top with Trotter errors in DQS of collective spin systems. Our approach
of harnessing the available knowledge on the kicked top, which is a paradigmatic
model of single-body quantum chaos, is complementary to the numerical study of
the connection between Trotter errors in DQS of generic many-body systems and
many-body quantum chaos presented in Ref.~\cite{Heyl2019}.

\sect{Methods}
\subsect{Higher-order Trotterization} In this work, we exploit the formal
equivalence of the dynamics of the kicked top which is determined by the Floquet
operator given in Eq.~\eqref{eq:U-kicked-top}, and the Trotterized dynamics in
DQS of collective spin systems as described by
Eq.~\eqref{eq:Suzuki-Trotter}. For a given target Hamiltonian $H = H_x + H_z$
which is composed of two non-commuting pieces $H_{x,z}$, Trotterization is not
unique. In the following, we briefly discuss different Trotterization schemes,
and how they affect our conclusions concerning Trotter errors and quantum chaos.

Trotterization of the unitary time evolution operator $U(t) = \e^{-\i H t}$ with
$H = H_x + H_z$ is carried out in two steps: First, the simulation time $t$ is
divided into $n$ Trotter steps of duration $\tau = t/n$,
\begin{equation}
  U(t) = \e^{-\i H t} = \left( \e^{-\i H t/n} \right)^n.
\end{equation}
Second, within a Trotter step of size $\tau$, the exponential $\e^{-\i H \tau}$
of the Hamiltonian $H$ is approximated by a product of exponentials of $H_x$ and
$H_z$ alone. This approximation introduces an error of order $\tau^m$, where $m$
can be increased by working with more elaborate
factorizations~\cite{Wiebe2010}. Equation~\eqref{eq:Suzuki-Trotter} corresponds
to the lowest-order decomposition with $m = 1$, and a first improvement can be
obtained by symmetrization,
\begin{align}
  \label{eq:first-order-Trotter}
  \e^{-\i H \tau} & = \e^{-\i H_x \tau} \e^{-\i H_z \tau} +
  O(\tau), \\
  \label{eq:second-order-Trotter}
  \e^{-\i H \tau} & = \e^{-\i H_x \tau/2} \e^{-\i H_z \tau} \e^{-\i H_x \tau/2} +
  O(\tau^2).
\end{align}
We emphasize that the given error bounds $O(\tau^m)$ pertain to the full time
evolution operator. In contrast, as shown in this work for collective spin
systems and in Ref.~\cite{Heyl2019} for a generic many-body system, Trotter
errors of few-body observables remain bounded even at long times, i.e., after
repeated application of the evolution operators in
Eqs.~\eqref{eq:first-order-Trotter} and~\eqref{eq:second-order-Trotter}
corresponding to elementary Trotter steps. However, it is an interesting
question for future studies how the threshold behavior in Trotter errors of
few-body observables and the quantum chaos properties discussed in Results are
affected by higher-order Trotterization. Notably, already going from the
first-order formula Eq.~\eqref{eq:first-order-Trotter} to the second-order
formula Eq.~\eqref{eq:second-order-Trotter} leads to a significant qualitative
difference in the PR Eq.~\eqref{equ:PR} shown in
Fig.~\ref{fig:signatures-chaos}\textbf{a}. This is discussed in detail in
Results. \\

\subsect{RMT predictions for the IPR, observables and OTOCs} For large Trotter
step sizes $\tau$, the Floquet operator Eq.~\eqref{eq:U-kicked-top} is faithful
to the COE of RMT. This is demonstrated in Results, where we consider the
participation ratio PR and eigenphase-spacing statistics,
cf.~Fig.~\ref{fig:signatures-chaos}. However, also the results for the long-time
averages of magnetization error and the simulation accuracy presented in
Results, as well as steady-state saturation of OTOCs, are indicative of the RMT
properties of the Floquet operator. Here, we derive RMT predictions for these
quantities.

RMT predictions for the IPR defined in Eq.~\eqref{eq:IPR} can be obtained by
averaging the overlap $\abs{\braket{\psi_0 | \phi_m}}^4$ over the distribution
of eigenvectors $\ket{\phi_m}$ of random matrices drawn from the appropriate
ensemble. In particular, the components of eigenvectors of random matrices
sampled from the CUE are distributed uniformly on the unit sphere in
$\C^{\mathcal{D}}$, while for the COE the components of eigenvectors can be
assumed real and thus lie on the the unit sphere in
$\R^{\mathcal{D}}$~\cite{Haake2010}. Averages over these eigenvector
distributions can be performed by explicitly parameterizing unit-norm
eigenvectors in terms of spherical coordinates and carrying out the resulting
integrals as discussed for the CUE in Ref.~\cite{PhysRev.132.948}. These averages
yield the same result for all states $\ket{\phi_m}$ with
$m = 1, \dotsc, \mathcal{D}$ and also do not depend on the reference state
$\ket{\psi_0}$. We note that for the COE, $\ket{\psi_0}$ should be chosen to
have real components. This is indeed the case for the eigenstates $\ket{\psi_n}$
of the target Hamiltonian which we consider in Eq.~\eqref{equ:PR}, if they are
expanded in the basis of eigenstates of $J_z$. We find
\begin{align}
  \label{eq:CUE-PR}
  \mathrm{IPR}_{\mathrm{CUE}} & = \frac{2}{\mathcal{D} + 1}, \\ \label{eq:COE-PR}
  \mathrm{IPR}_{\mathrm{COE}} & = \frac{3}{\mathcal{D} + 2}.
\end{align}
The RMT predictions for the PR as defined in Eq.~\eqref{equ:PR} follow from the
above results for the IPR as $\mathrm{PR} = 1/(\mathcal{D} \, \mathrm{IPR})$,
and are shown in Fig.~\ref{fig:signatures-chaos}\textbf{a}.

With regard to the simulation accuracy $Q_E$ defined in Eq.~\eqref{eq:QE}, the
long-time average value of $\overline{Q_E} = 1$ which is reached in the chaotic
regime as shown in Fig.~\ref{fig:Trotter-errors}\textbf{b,e}, indicates that the
temporal average of the energy
$E_{\tau}(t) = \left\langle H(t) \right\rangle_{\tau}$ is equivalent to taking
the expectation value of $H$ in an infinite-temperature state, i.e.,
$E_{T = \infty} = \tr(H)/\mathcal{D}$. As we show now, the same result is
predicted by RMT. To this end, we replace in Eq.~\eqref{eq:QE} the $n_t$-th
power of Floquet operator, $U_{\tau}^{n_t}$ with $n_t = \lfloor t/\tau \rfloor$,
by $Q^{\dagger} U Q$ where $Q$ is given in
Eq.~\eqref{eq:first-to-second-order-Trotter} and $U$ is a random unitary matrix
from the COE. As discussed in Results, the COE is comprised of symmetric
matrices, and the unitary transformation $Q$ acts to symmetrize the Floquet
operator $U_{\tau}$. Then, the average over $U \in \mathrm{COE}$, which we
denote by $[\dotsm]_{\mathrm{COE}}$ in the following, yields~\cite{Brouwer1996}
\begin{equation}
  \begin{split}
    E_{\mathrm{COE}} & = \left[ \braket{\psi_0 | Q^{\dagger} U^{\dagger} Q H
        Q^{\dagger} U Q | \psi_0} \right]_{\mathrm{COE}} \\ & =
    \frac{1}{\mathcal{D} + 1} \left( \tr(H) + \braket{\psi_0 | Q^{\dagger}
        \tilde{H}^T Q | \psi_0} \right) \\ & \sim \frac{1}{\mathcal{D}} \tr(H) =
    E_{T = \infty} \quad \text{for} \quad \mathcal{D} \to \infty,
  \end{split}
\end{equation}
where $\tilde{H} = Q H Q^{\dagger}$. Hence, we obtain the RMT prediction for the
simulation accuracy in the chaotic regime,
$Q_{\mathrm{COE}} = (E_{\mathrm{COE}} - E_0)/(E_{T = \infty} - E_0) \sim 1$ for
$\mathcal{D} \to \infty$.

Similarly, Figs.~\ref{fig:otocs}\textbf{a,b} show that at late times the OTOC
$C(t)$ Eq.~\eqref{eq:otoc} approaches a value $C_{\mathrm{COE}}$ which as we
explain in the following can also be obtained by taking an average over the
COE. First, we consider the COE average of the OTOC $F(t)$ defined as
\begin{align}
  \label{eq:F-otoc}
  F(t) = \frac{1}{\mathcal{D}} \tr \! \left( W(t)^{\dagger} V^{\dagger} W(t) V \right),
\end{align}
where $W(t) = U_{\tau}^{n_t \dagger} W U_{\tau}^{n_t}$ and
$n_t = \lfloor t/\tau \rfloor$. We assume $\tr(W) = 0$ and
$V^{\dagger} = V^{-1}$, which is satisfied for $W = S_z$ and
$V = \e^{-\i \varphi S_z}$ shown in Fig.~\ref{fig:otocs}. As above, we replace
$U_{\tau}^{n_t}$ by $Q^{\dagger} U Q$, and by taking the average over
$U \in \mathrm{COE}$ we find
\begin{multline}
  \label{eq:otoc-coe}
  F_{\mathrm{COE}} = \frac{1}{\mathcal{N}_{\mathcal{D}}} \left[ \left(
      \mathcal{D}+2 \right) \left(
      \tr( \tilde{V} \tilde{W}^* \tilde{V}^{\dagger} \tilde{W}^T ) \right. \right. \\
  \left. + \tr( \tilde{V} \tilde{W}^T \tilde{V}^{\dagger} \tilde{W}^* ) + \bigl|
    \tr( \tilde{V}) \bigr|^2 \tr( \tilde{W} \tilde{W} ^{\dagger} ) \right) \\ -
  \left( \mathcal{D}+4 \right) \tr( \tilde{W}^2 ) - \bigl| \tr( \tilde{V}
  \tilde{W}^T) \bigr|^2 - \bigl| \tr( \tilde{V} \tilde{W}^*) \bigr|^2 \\
  \left. - \left( \mathcal{D} + 1 \right) \left( \tr( \tilde{V} ) \tr(
      \tilde{V}^{\dagger} \tilde{W}^T \tilde{W}^* ) + \tr( \tilde{V}^{\dagger} )
      \tr( \tilde{V} \tilde{W}^* \tilde{W}^T ) \right) \right],
\end{multline}
where
$\mathcal{N}_{\mathcal{D}} = \mathcal{D}^2 \left( \mathcal{D}+1 \right) \left(
  \mathcal{D}+3 \right),$
$\tilde{W} = Q W Q^{\dagger}$, and $\tilde{V} = Q V Q^{\dagger}$. 
To relate $F(t)$ Eq.~\eqref{eq:F-otoc} to the squared commutator in
Eq.~\eqref{eq:otoc}, we note that for $W = S_z$ we obtain
$\tr(W^2)/\mathcal{D} = S \left( S + 1 \right) \! /3$, which yields
\begin{equation}
  \label{eq:C-F-otoc}
  C(t) = 2 \left( \frac{S \left( S + 1 \right)}{3} - \Re( F(t) ) \right).
\end{equation}
This relation holds at all times $t$ and, in particular, it allows us to express
the COE average $C_{\mathrm{COE}}$ in terms of $F_{\mathrm{COE}}$ given in
Eq.~\eqref{eq:otoc-coe}. Moreover, the infinite-time average $\overline{C}$
shown in Fig.~\ref{fig:otocs} can be obtained in terms of the corresponding
average $\overline{F}$ of $F(t)$, which reads
\begin{multline}
  \label{eq:F-otoc-infinite-time-avg}
  \overline{F} = \frac{1}{\mathcal{D}} \sum_{m, n = 1}^{\mathcal{D}} \frac{1}{1
    + \delta_{m, n}} \left( W_{mm}^{} V^{\dagger}_{mn} W_{nn}^{} V_{nm}^{}
  \right. \\ \left. + W_{mn}^{} V^{\dagger}_{nn} W_{nm}^{} V_{mm}^{} \right),
\end{multline}
where $W_{n m} = \braket{\phi_n | W | \phi_m}$ etc.\ and the vectors
$\ket{\phi_n}$ with $n = 1, \dotsc, \mathcal{D}$ form an eigenbasis of the
Floquet operator. Figure~\ref{fig:otocs} shows that the RMT expression
$C_{\mathrm{COE}}$ of the OTOC agrees with the temporal average obtained from
Eq.~\eqref{eq:F-otoc-infinite-time-avg} in the chaotic regime and beyond the
Ehrenfest time $t_{\mathrm{E}}$. \\

\subsect{The semiclassical limit of the kicked top} As pointed out above, in the
kicked top, the role of an effective Planck constant is played by the inverse
spin size, $\hbar_{\mathrm{eff}} = 1/S$.  Hence, if the kicked top is realized
as the collective spin of a system of $N$ spin-\nicefrac{1}{2} such that
$S = N/2$, the thermodynamic limit $N \to \infty$ coincides with the
semiclassical limit $\hbar_{\mathrm{eff}} \to 0$. In this limit, the spin $S$
obeys stroboscopic evolution equations that can be obtained from the Heisenberg
equations of motion for the spin components $S_{x, y, z}$ by introducing
rescaled variables,
\begin{equation}
  \label{eqn:CLvars}
  X = S_x/S, \qquad Y = S_y/S, \qquad Z = S_z/S,
\end{equation}
and subsequently taking the limit $S \to \infty$. In this limit, the commutators
$[X, Y]$ etc.\ vanish, and $X$, $Y$, and $Z$ become effectively classical
variables.

The stroboscopic evolution equations of the spin operators in the Heisenberg
picture are determined by
\begin{equation}
  \label{eqn:heissenberg}
  S_i(t+\tau) = U_\tau^{\dagger} S_i(t) U_\tau \quad \text{for} \quad i \in \{x,y,z\}.
\end{equation}
To evaluate the right-hand side of this equation, we use relations such
as~\cite{Kitagawa1993}
\begin{equation}
  \label{eqn:ueda}
  \e^{\i \tau F(S_z)} S_+ \e^{-\i \tau F(S_z)} = S_+ \e^{\i \tau f(S_z)},
\end{equation}
where $f(S_z) = F(S_z + 1) - F(S_z)$, and which holds for any analytic
function $F$ and pairs of operators $S_z$ and $S_+ = S_x + \i S_y$ that obey
$[S_z, S_+] = S_+$. We thus arrive at
\begin{widetext}
  \begin{equation}
    \label{eq:semiclassical-eoms}
    \begin{split}
      X(t+\tau) & = \Re \! \left( \e^{\i \tau h_z} \left[ X(t) + \i \Re
          \left((Y(t) + \i Z(t)) \e^{\i \tau h_x} \e^{\i J_x X(t) \tau} \right)
        \right] \e^{ \frac{J_z \tau}{2} \Im \left(
            (Y(t) + \i Z(t)) \e^{\i \tau
              h_x} \e^{\i J_x X(t) \tau} \right)} \right), \\
      Y(t+\tau) &= \Im \! \left( \e^{\i \tau h_z} \left[ X(t) + \i \Re
          \left((Y(t) + \i Z(t)) \e^{\i \tau h_x} \e^{\i J_x X(t) \tau} \right)
        \right] \e^{\frac{J_z \tau}{2} \Im \left( (Y(t) + \i Z(t)) \e^{\i \tau
              h_x} \e^{\i J_x X(t) \tau} \right)}
      \right), \\
      Z(t+\tau) & = \Im \! \left( (Y(t) + \i Z(t)) \e^{\i \tau h_x} \e^{\i J_x X(t)
          \tau} \right).
    \end{split}
  \end{equation}
\end{widetext}
To obtain the simulation accuracy Eq.~\eqref{eq:QE} in the limit $S \to \infty$,
we iterate the semiclassical evolution equations~\eqref{eq:semiclassical-eoms}
and insert the resulting values of the spin components as per
Eq.~\eqref{eqn:CLvars} in the target Hamiltonian $H = H_x + H_z$, which we
interpret as a function of the classical variables $S_{x, y, z}$. This yields $E_{\tau}(t)$. The
energy $E_{T = \infty}$ at infinite temperature is obtained by averaging $H$
over the classical phase space, i.e., a sphere of radius $S$. Finally, $E_0$ is
the energy of the initial spin configuration.

The magnetization error Eq.~\eqref{eq:magnetization-error} can be obtained
similarly: In the semiclassical limit,
$\left\langle S_z(t) \right\rangle_{\tau}$ in Eq.~\eqref{eq:magnetization-error}
can be replaced by $S Z(t)$, where $Z(t)$ is evolved with
Eq.~\eqref{eq:semiclassical-eoms}; To find $\left\langle S_z(t) \right\rangle$,
we again interpret $H$ as a classical Hamiltonian and integrate the
corresponding Hamiltonian evolution equations up to the time $t$.

\sect{Acknowledgments} We thank B. Vermersch, M. Baranov, and M. Bukov for
helpful discussions. Work in Innsbruck is supported by the European Research Council
(ERC) Synergy Grant UQUAM and the SFB FoQuS (FWF Project No. F4016-N23). MH
acknowledges support by the Deutsche Forschungsgemeinschaft (DFG) via the
Gottfried Wilhelm Leibniz Prize program. PH acknowledges support by the DFG
Collaborative Research Centre SFB 1225 (ISOQUANT), the ERC Advanced Grant
EntangleGen (Project-ID 694561), and the ERC Starting Grant StrEnQTh. Numerical
simulations were realized with
QuTiP~\cite{Johansson20131234}. \\


\end{document}


\title{Supplementary Information: Digital Quantum Simulation, Trotter Errors, and Quantum Chaos of the
  Kicked Top}

\author{Lukas M. Sieberer}

\email{lukas.sieberer@uibk.ac.at}

\affiliation{Center for Quantum Physics, University of Innsbruck, 6020
  Innsbruck, Austria}

\affiliation{Institute for Quantum Optics and Quantum Information of the
  Austrian Academy of Sciences, A-6020 Innsbruck, Austria}

\author{Tobias Olsacher}

\affiliation{Center for Quantum Physics, University of Innsbruck, 6020
  Innsbruck, Austria}

\author{Andreas Elben}

\affiliation{Center for Quantum Physics, University of Innsbruck, 6020
  Innsbruck, Austria}

\affiliation{Institute for Quantum Optics and Quantum Information of the
  Austrian Academy of Sciences, A-6020 Innsbruck, Austria}

\author{Markus Heyl}

\affiliation{Max Planck Institute for the Physics of Complex Systems,
  N\"othnitzer Str.~38, 01187 Dresden, Germany}

\author{Philipp Hauke}

\affiliation{Kirchhoff-Institute for Physics, Heidelberg University, 69120
  Heidelberg, Germany}

\affiliation{Institute for Theoretical Physics, Heidelberg University, 69120
  Heidelberg, Germany}

\author{Fritz Haake}

\affiliation{Faculty of Physics, University of Duisburg-Essen, 47048 Duisburg, Germany}

\author{Peter Zoller}

\affiliation{Center for Quantum Physics, University of Innsbruck, 6020
  Innsbruck, Austria}

\affiliation{Institute for Quantum Optics and Quantum Information of the
  Austrian Academy of Sciences, A-6020 Innsbruck, Austria}

\date{\today}

\maketitle

\section{Perturbation Theory}
\label{sec:PertTheory}

In the regular regime of Trotter step sizes below $\tau_{*}$, corrections to the
ideal long-time averages of observables due to Trotterization can be obtained by
treating the difference between the effective Hamiltonian $H_{\tau}$ obtained
from the FM expansion and the target Hamiltonian $H$ in time-dependent
perturbation theory as described in Ref.~\cite{Heyl2019}. Here, we generalize
the discussion of Ref.~\cite{Heyl2019}, which assumed a fixed initial state
$\ket{\psi_0}$ polarized along the $z$-axis, to an arbitrary choice of
$\ket{\psi_0}$.

The starting point is the FM expansion of the effective Hamiltonian $H_{\tau}$
up to second order in $\tau$ (see Ref.~\cite{Bandyopadhyay2015} in the context
of the kicked top)
\begin{equation}
  \label{eqn:FloquetHamil}
  H_{\tau} = H + \tau C_1 + \tau^2 C_2 = H + \frac{\i \tau}{2 } [H_x,H_z]
  -\frac{\tau^2}{12 ^2} [H_x - H_z, [H_x, H_z]], 
\end{equation}
where the operator $V = \tau C_1 + \tau^2 C_2$ acts as a time-independent
correction to the target Hamiltonian $H = H_x + H_z$. The
effective Hamiltonian gives rise to a time evolution according to the
Schr\"odinger equation
\begin{equation}
\i \partial_t \ket{\psi} = H_{\tau} \ket{\psi}
\end{equation}
which we transform into an interaction picture according to
\begin{equation}
\ket{\psi_I(t)} = \e^{\i H t} \ket{\psi}, \qquad V_I = \e^{\i H t} V \e^{-\i H t}.
\end{equation}
We obtain the time-evolution operator from time $t_0$ to $t$ as
\begin{equation}
W(t,t_0) = \mathsf{T} \e^{-\i \int_{t_0}^t V_I(s) \; \mathrm{d} s}
\end{equation}
where $\mathsf{T}$ denotes a time-ordered exponential. We determine
$W(t) = W(t,0)$ ($t_0=0$) up to second order in the Trotter step $\tau$.

\subsection{Corrections to the energy}
We are interested in an expansion of the quantity
\begin{equation}
\begin{split}
  \Delta E(t) & = \left\langle H(t) \right\rangle_{\tau} - E_0 = \langle \psi_0
  \vert \e^{\i H_{\tau} t} H \e^{-\i H_{\tau} t} \vert \psi_0 \rangle - \langle
  \psi_0 \vert H \vert \psi_0 \rangle \\
  & = \tau \left(\langle \psi_0 \vert C_1 \vert \psi_0 \rangle - \langle \psi_0
    \vert \e^{\i H_{\tau} t} C_1 \e^{-\i H_{\tau} t} \vert \psi_0 \rangle \right) +
  \tau^2 \left(\langle \psi_0 \vert C_2 \vert \psi_0 \rangle - \langle \psi_0
    \vert \e^{\i H_{\tau} t} C_2 \e^{-\i H_{\tau} t} \vert \psi_0 \rangle \right),
\end{split}
\end{equation}
where we used Eq.~\eqref{eqn:FloquetHamil}, in the Trotter step $\tau$ for an
arbitrary initial state $\ket{\psi_0}$. To this end, we replace
$\e^{-\i H_{\tau} t}$ by $W(t)$ which yields the time dependent correction up to
second order in the Trotter step $\tau$:
\begin{multline}
  \Delta E(t) = \tau \langle \psi_0 \vert C_1 - C_{1, I}(t) \vert \psi_0 \rangle +
  \frac{\tau^2}{4} \langle \psi_0 \vert \e^{\i H t} \left[H_z, [H_x,H_z] \right]
  \e^{-\i H t} \vert \psi_0 \rangle \\ + \frac{\i \tau^2}{2} \langle \psi_0 \vert
  [H_z, C_{1, I}(t)] \vert \psi_0 \rangle + \tau^2 \langle \psi_0 \vert C_2 -
  C_{2, I}(t) \vert \psi_0 \rangle + O(\tau^3).
\end{multline}
The long-time averages of the above expectation values are determined in the
diagonal ensemble~\cite{Rigol2008},
\begin{equation}
  \label{eqn:DiagEns}
  \overline{\langle \psi_0 \vert O \vert \psi_0 \rangle} = \lim_{t\rightarrow
    \infty} \frac{1}{t} \int_0^t \langle \psi_0 \vert O(t) \vert \psi_0 \rangle
  \, d t = \sum_{\lambda \in \sigma(H)} \vert \langle \psi_0 \vert \lambda \rangle \vert^2 \langle \lambda \vert O \vert \lambda \rangle
\end{equation}
where the states $\ket{\lambda}$ form a basis of eigenstates of the target
Hamiltonian, $H \ket{\lambda} = \lambda \ket{\lambda}$. This simplifies the
correction to
\begin{multline}
  \overline{\Delta E} = \tau \langle \psi_0 \vert C_1 \vert \psi_0 \rangle +
  +\tau^2 \langle \psi_0 \vert C_2 \vert \psi_0 \rangle - \frac{\tau^2}{4}
  \sum_{\lambda \in \sigma(H)} \vert \langle \psi_0 \vert \lambda \rangle
  \vert^2 \langle \lambda \vert \left[ H_z, [H_x,H_z] \right] \vert \lambda
  \rangle \\ - \tau^2 \sum_{\lambda \in \sigma(H)} \Im \left( \langle \lambda
    \vert \psi_0 \rangle \langle \psi_0 \vert H_z \vert \lambda \rangle \langle
    \lambda \vert C_1 \vert \lambda \rangle \right) - \frac{\tau^2}{12}
  \sum_{\lambda \in \sigma(H)} \vert \langle \psi_0 \vert \lambda \rangle
  \vert^2 \langle \lambda \vert \left[ H_x-H_z, [H_x,H_z] \right] \vert \lambda
  \rangle + O(\tau^3).
\end{multline}
This expression yields the dotted lines in Fig.~1\textbf{e} of the main text (MT).

\subsection{Corrections to arbitrary observables}

In the same way as before we derive an expansion up to second order in the
Trotter step $\tau$ for the correction to an arbitrary observable $A$ starting
from an initial state $\vert \psi_0 \rangle$. After taking long-time averages in
the diagonal ensemble we obtain the result
\begin{multline}  
  \overline{A} =
  \sum_{\lambda \in \sigma(H)} \vert \langle \psi_0 \vert \lambda \rangle \vert^2 \langle \lambda \vert A \vert \lambda \rangle - \tau \Im \left( \sum_{\lambda \in \sigma(H)} \vert \langle \lambda \vert \psi_0 \rangle \vert^2 \langle \lambda \vert H_z A \vert \lambda \rangle - \sum_{\lambda \in \sigma(H)} \langle \lambda \vert \psi_0 \rangle \langle \psi_0 \vert H_z \vert \lambda \rangle \langle \lambda \vert A \vert \lambda \rangle \right) \\
  + \frac{\tau^2}{4} \left( \sum_{\lambda \in \sigma(H)} \vert \langle \lambda \vert \psi_0 \rangle \vert^2 \langle \lambda \vert H_z A H_z \vert \lambda \rangle + \sum_{\lambda \in \sigma(H)} \vert \langle \psi_0 \vert H_z \vert \lambda \rangle \vert^2 \langle \lambda \vert A \vert \lambda \rangle \right) - \frac{\tau^2}{2} \Re \left(\sum_{\lambda \in \sigma(H)} \langle \lambda \vert H_z \vert \psi_0 \rangle \langle \psi_0 \vert \lambda \rangle \langle \lambda \vert H_z A \vert \lambda \rangle \right) \\
  + \frac{\tau^2}{6} \Re \Bigg( \sum_{\lambda_1 \neq \lambda_2} \Big( \langle
  \psi_0 \vert \lambda_2 \rangle \langle \lambda_2 \vert A \vert \lambda_1
  \rangle - \langle \psi_0 \vert \lambda_1 \rangle \langle \lambda_1 \vert A
  \vert \lambda_1 \rangle \Big) \langle \lambda_2 \vert \psi_0 \rangle \langle
  \lambda_1 \vert \left[ H_x-H_z, [H_x, H_z] \right] \vert \lambda_2 \rangle
  \frac{1}{\lambda_1-\lambda_2} \Bigg) \\ - \frac{\tau^2}{2} \Re \Bigg(
  \sum_{\lambda_1 \neq \lambda_2} \Big( \langle \psi_0 \vert \lambda_2 \rangle
  \langle \lambda_2 \vert A \vert \lambda_1 \rangle - \langle \psi_0 \vert
  \lambda_1 \rangle \langle \lambda_1 \vert A \vert \lambda_1 \rangle \Big)
  \langle \lambda_2 \vert \psi_0 \rangle \langle \lambda_1 \vert [H_x, H_z] H_z
  \vert \lambda_2 \rangle \frac{1}{\lambda_1-\lambda_2} \Bigg) \\ +
  \frac{\tau^2}{2} \Re \left( \sum_{\lambda \in \sigma(H)} \langle \psi_0 \vert
    \lambda \rangle \langle \lambda \vert H_z \vert \psi_0 \rangle \langle
    \lambda \vert A H_z \vert \lambda \rangle \right) - \frac{\tau^2}{2} \Re
  \left( \sum_{\lambda \in \sigma(H)} \langle \psi_0 \vert \lambda \rangle
    \langle \lambda \vert A \vert \lambda \rangle \langle \lambda \vert H_z^2
    \vert \psi_0 \rangle \right) + O(\tau^3).
\end{multline}

\section{Measurement of OTOCs using statistical correlations of randomized
  measurements}
\label{sec:meas-otocs-using}

The onset of chaos and thus the breakdown of DQS is diagnosed by the OTOC
defined in Eq.~(17) of MT. In the following, we describe here how the OTOC can
be measured in experiments.

The measurement of the OTOC is very challenging due to the alternation between
operators at time zero and a finite time $t$. Protocols to measure $C(t)$ (or
$F(t)$ defined in Eq.~(24) of MT) that have been proposed in the literature thus
involve backward time evolution~\cite{Swingle2016, Zhu2016, Bohrdt2017,
  Shen2017, Yoshida2018} or the coupling to auxiliary quantum
systems~\cite{Yao2016}, and the first measurements have been implemented only
recently~\cite{Li2017, Garttner2017, Landsman2018}. Here we propose to measure
the OTOC by adapting a recently developed measurement
protocol~\cite{Vermersch2018a} which does not require the above
ingredients. Instead, it is an example of measurement protocols which are based
on statistical correlations between randomized measurements. This concept was
first devised for R\'enyi entropies~\cite{VanEnk2012, Elben2017, Vermersch2018},
and was recently demonstrated in a trapped-ion quantum
simulator~\cite{Brydges2018} and in NMR~\cite{Nie2019}. The crucial ingredient
in this class of measurement protocols are random unitary transformations which
form (approximate) unitary two-designs. Thus, we first comment on the generation
of such random unitaries in collective spin models and subsequently discuss how
to use them to measure the OTOC.

\subsection{Generation of random unitaries}
\label{sec:gener-rand-unit}

In systems realizing collective spin models, the same tools that are used to
engineer Hamiltonians of the form given in Eq.~(7) of MT can be applied to
generate random unitaries from approximate unitary two-designs. Unitary
two-designs are ensembles that approximate the circular unitary ensemble (CUE)
up to $2$nd order correlations~\cite{Roberts2017}. We adapt ideas from
Ref.~\cite{Elben2017, Vermersch2018} to create such random unitaries from a
series of random quenches as a random time evolution operator $u$,
\begin{equation}
  u=u_\eta \cdots u_1 \qquad \text{with} \qquad u_\alpha= \e^{-\i H^\alpha_z \tau}\e^{-\i H^\alpha_x\tau} 
\end{equation}
where the Hamlitonians $H_{\mu}^{\alpha}$ take the same form as in Eq.~(7) of
MT,
\begin{equation}
H_\mu^\alpha =  h_\mu^\alpha S_\mu + J_\mu S_\mu^2/ (2S + 1).
\end{equation}
For each quench, parameters $h_{\mu}^{\alpha}$ and $J_{\mu}^{\alpha}$ are chosen
independently from uniform distributions $h_\mu^\alpha \in [0.5/\tau,1.5/\tau]$,
$J_\mu^\alpha \in [5.5/\tau, 6.5/\tau]$ such that the kicked top which would be
realized by periodic repetition of any of the $u_\alpha$ is deep in the chaotic
regime~\cite{Haake1987}.

\begin{figure}
  \centering
  \includegraphics{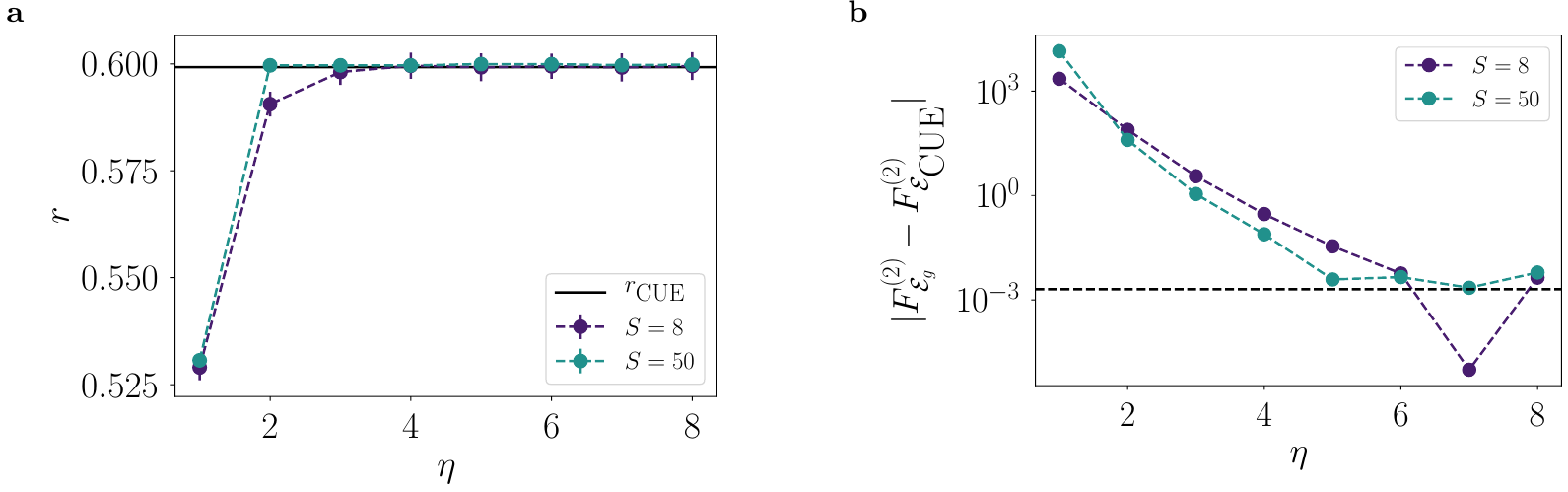}
  \caption{Generation of approximate two-designs. Convergence of $N_U=500$
    generated random unitaries to an approximate two-design as a function of the
    number of random quenches $\eta$ for different spin sizes $S$. \textbf{a} Average
    adjacent phase spacing ratio. The black line corresponds to the CUE
    value. \textbf{b} Difference of the frame potential $F_{\mathcal{E}_g}$ of the
    generated unitaries to the frame potential
    $F_{\mathcal{E}_\text{CUE}}=k!+d^4/N_U$ of an ensemble of same size
    $|\mathcal{E}_g|=|\mathcal{E}_{\mathrm{CUE}}|=500$ which is drawn from the
    CUE. The black dashed line is the statistical error threshold
    $\sim 1/|\mathcal{E}_g|$.}
  \label{Fig1:Unitary}
\end{figure}

To test whether the ensemble of thusly created random unitaries indeed forms a
unitary two-design, we consider first their phase spacing statistics as
characterized by the ratios of adjacent phase spacing~\cite{DAlessio2013}. As
displayed in Fig.~\ref{Fig1:Unitary}, we observe a very rapid convergence of the
mean value $\langle r \rangle$ with the number of quenches $\eta$ towards the
CUE value $r_{\mathrm{CUE}} = 0.5996(1)$. Second, to test not only eigenvalues
but also the eigenvectors of the created unitaries, we consider the second frame
potential~\cite{Roberts2017}, which is defined as
\begin{equation}
  F^{(2)}_{\mathcal{E}} = \frac{1}{|\mathcal{E}|} \sum_{U,V \in \mathcal{E}}^{}
  \abs{\tr ( U^{\dagger} V )}^4
  \label{eq:frame}
\end{equation}
where $\mathcal{E}$ is an ensemble of unitary matrices.  As shown in
Ref.~\cite{Roberts2017}, for $\mathcal{E}$ being a unitary two-design, the frame
potential takes a minimal value $F^{(2)}_{\text{CUE}}=2!$, and
$F^{(2)}_{\mathcal{E}}-F^{(2)}_{\text{CUE}}$ represents a fine-grained measure
of distance between an ensemble $\mathcal{E}$ and a unitary two-design. However,
we note that for small ensembles $|\mathcal{E}|\ll \mathcal{D}^4$, the diagonal
contribution $\mathcal{D}^4/|\mathcal{E}|$ in the sum Eq.~\eqref{eq:frame}
dominates, irrespective of the properties of $\mathcal{E}$. Thus, to test
numerically using a finite sample size whether the generated random unitaries
are sampled from an approximate unitary two-design, we consider the quantity
$F^{(2)}_{\mathcal{E}}-F^{(2)}_{\mathcal{E}_\text{CUE}}$. Here,
$F^{(2)}_{\mathcal{E}_\text{CUE}}= 2! + \mathcal{D}^4/|\mathcal{E}_\text{CUE}|$
is the average frame potential of a finite number $|\mathcal{E}_\text{CUE}| $ of
unitaries sampled from the CUE (see also Ref.~\cite{Cotler2017}).  In
Fig.~\ref{Fig1:Unitary},
$F^{(2)}_{\mathcal{E}}-F^{(2)}_{\mathcal{E}_\text{CUE}}$ is displayed as
function of number of quenches $\eta$, for a sample size
$|\mathcal{E}_\text{CUE}| = |\mathcal{E}_g|=500$.  We observe rapid convergence
of $F^{(2)}_{\mathcal{E}_g}-F^{(2)}_{\text{CUE}}$ towards the statistical error
threshold $\sim 1/|\mathcal{E}_g|$, signaling that the generated random
unitaries are indeed sampled from an (approximate) unitary two-design.

\subsection{Protocol to measure the OTOC}
\label{sec:measurement-otoc}

\begin{figure}
  \centering
  \includegraphics[width=0.4\textwidth]{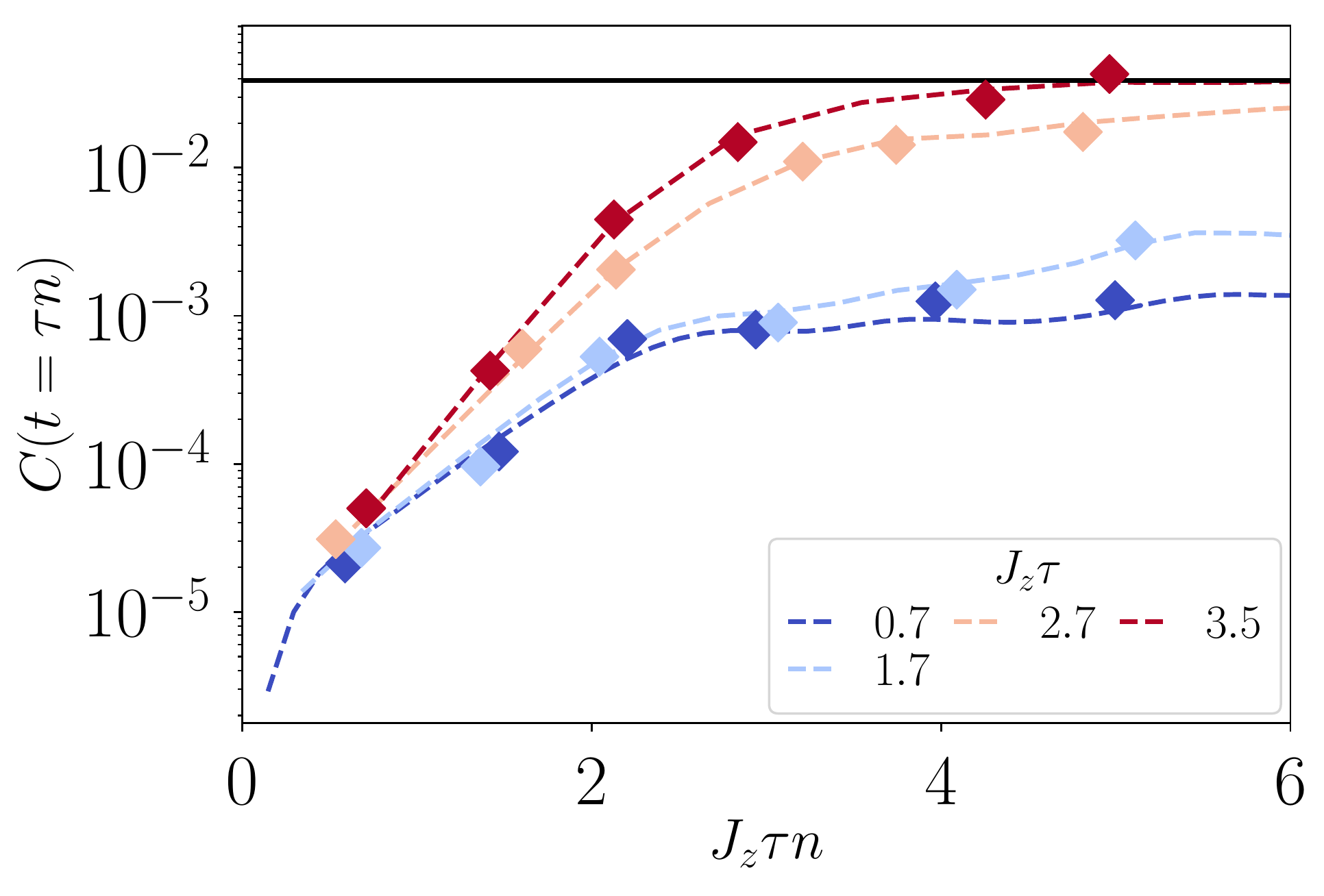}
  \caption{\textit{Measurement of the OTOC.} Time evolution of the exact (solid
    lines) and ``measured'' (dots) squared commutator
    $C(t)=2 \left( S \left(S+1 \right) \! /3 - \Re(F(t) \right)$ for a system
    with collective spin $S=64$.  Parameters are the same as in Fig.~2 of the
    main text. The ``measurements'' are simulated using $N_U=100$ random
    unitaries generated with $\eta=5$ random quenches (same parameters as in
    Fig.~\ref{Fig1:Unitary}). Projection noise is not included. The black line
    is the COE value.}
  \label{Fig2:MeasOTOC}
\end{figure}

The generated random unitaries, sampled from an approximate unitary two-design,
can be used to measure the OTOC according to the following
recipe~\cite{Vermersch2018a}: (i) To an arbitrary pure state $\ket{\psi_0}$,
apply a random unitary $u$ generated via random quenches (see
Appendix~\ref{sec:gener-rand-unit}), to prepare the state
$\ket{\psi_u} = u \ket{\psi_0}$. This randomized state serves as the input for
two independent experiments.  (ii.a) Evolve the state in time with $U(t)$ and
measure the expectation value
$\braket{ W(t) }_u=\braket{\psi_u| U^{\dagger}(t) W U(t) | \psi_u}$ of the
operator $W$. (ii.b) The state $\psi_u$ is perturbed by the operator $V$,
evolved in time and $W$ is measured, to obtain
$\braket{V^{\dagger} W(t) V }_u=\braket{\psi_u| V^{\dagger} U^{\dagger}(t) W
  U(t) V | \psi_u}$.
(iii) Repeat steps (i) and (ii) for a sample of random unitaries $u$ to estimate
the statistical correlations of the two expectation values obtained in (ii.a)
and (ii.b) across the unitary ensemble. As shown in Ref.~\cite{Vermersch2018a},
the OTOC can be expressed in terms of these statistical correlations as
\begin{equation}
  F(t)/F(0) = \frac{\overline{\langle W(t) \rangle_u \langle V^{\dagger} W(t) V
      \rangle_u}}{\left(\overline{\langle W(t) \rangle_u ^2 }\;\overline{\langle
        V^{\dagger}  W(t)  V\rangle_u ^2 }\right)^{{1}/{2}}},
\label{eq:Otilde}
\end{equation}
where $\overline{\vphantom{V}\cdots{}}$ denotes the ensemble average over random
unitaries $u$. This expression underpins the interpretation of $F(t)$ as a
measure of the butterfly effect in a quantum system: In analogy to rapidly
diverging classical chaotic trajectories, the expectation values in the
numerator can become decorrelated due to the perturbation $V$ of the initial
state. This is diagnosed by the decay of $F(t)$. In Fig.~\ref{Fig2:MeasOTOC},
data points correspond to simulated experiments using $100$ generated random
unitaries. The data demonstrates that the protocol outlined above can serve as
another experimental probe to delimit the region of applicability of DQS and the
onset of quantum chaos.

